\documentclass[
reprint, prd, aps, twocolumn, nofootinbib, preprintnumbers, superscriptaddress, balancelastpage, longbibliography,
 amsmath,amssymb,
 aps,
]{revtex4-2}

\newcommand{\dif}{\textnormal{\slshape d}}
\newcommand{\Fermi}{\textit{Fermi}}

\newcommand{\FermiLAT}{\textit{Fermi}-LAT }
\newcommand{\UMa}{Ursa Major III}

\newcommand{\beq}{\begin{equation}}
\newcommand{\eeq}{\end{equation}}

\newcommand{\mnras}{Monthly Notices of the Royal Astronomical Society}
\newcommand{\apjs}{The Astrophysical Journal Supplement Series}
\newcommand{\jcap}{Journal of Cosmology and Astroparticle Physics}

\newcommand\orcid[1]{\href{https://orcid.org/#1}{$\!$\includegraphics[scale=0.0045]{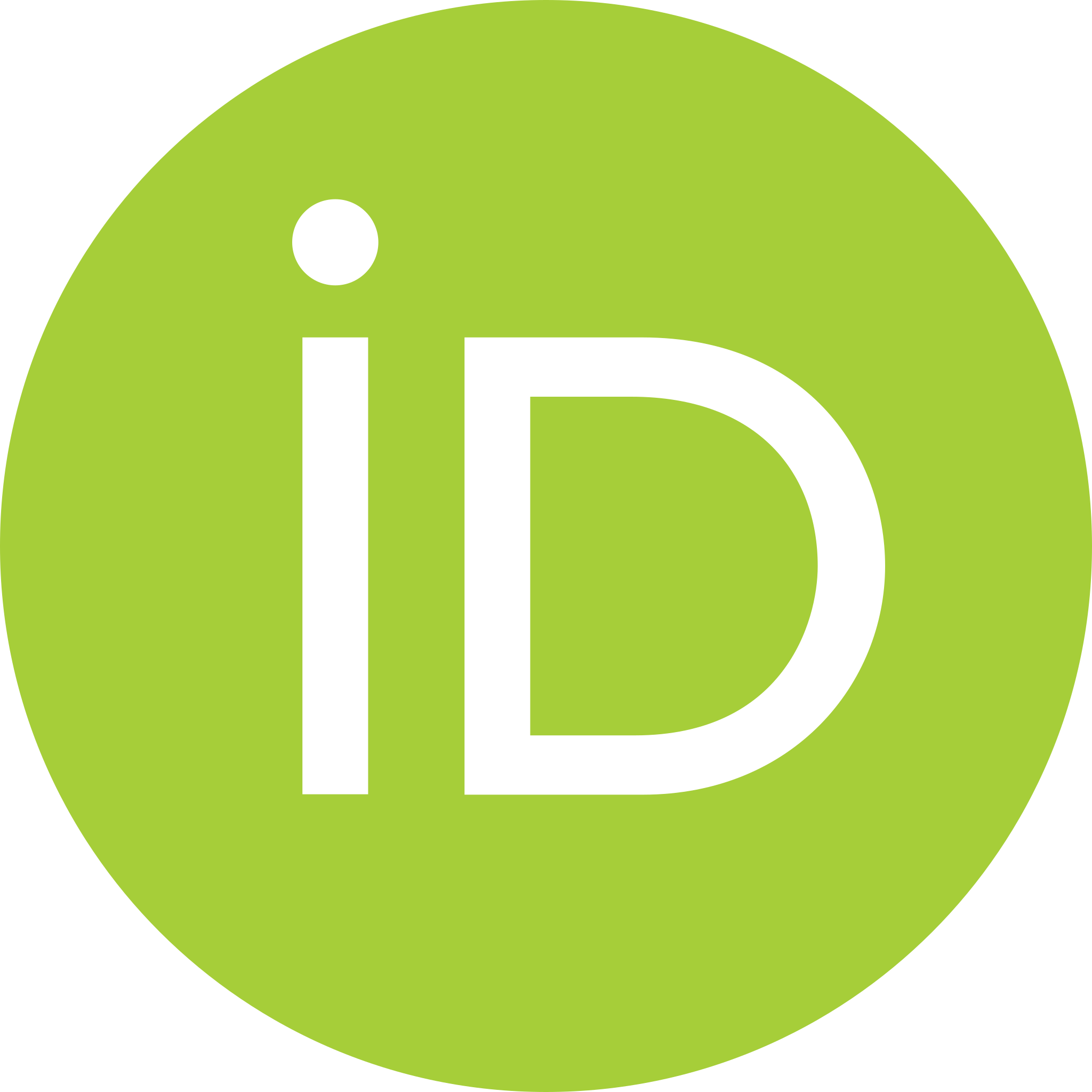} $\!\!$}}

\usepackage{graphicx}
\usepackage{dcolumn}
\usepackage{bm}
\usepackage{hyperref}
\usepackage{color}
\usepackage[dvipsnames]{xcolor}

\hypersetup{
     colorlinks   = true,
     citecolor    = cyan!75!Blue,
     urlcolor     = cyan!75!Blue,
     linkcolor    = cyan!75!Blue
}

\begin{document}


\title{Strong Constraints on Dark Matter Annihilation in Ursa Major III/UNIONS 1}


\author{Milena Crnogor\v{c}evi\'{c} \orcid{0000-0002-7604-1779}}\email{milena.crnogorcevic@fysik.su.se}
\author{Tim Linden \orcid{0000-0001-9888-0971}}\email{linden@fysik.su.se}
\affiliation{The Oskar Klein Centre, Department of Physics, Stockholm University, Stockholm 106 91, Sweden}

\date{\today}

\begin{abstract}
Very recent work has identified a new satellite galaxy, Ursa Major III/UNIONS I, which is the faintest such system ever observed. Dynamical considerations indicate that if the system is in equilibrium, it is likely to be highly dark matter dominated. This, in combination with its proximity, predicts that it may be the preeminent dwarf spheroidal galaxy target for dark matter indirect detection searches. We utilize 15 years of \Fermi-LAT data to search for $\gamma$-ray emission from Ursa Major III. Finding no excess, we set strong constraints on dark matter annihilation. Intriguingly, if the high J-factor of Ursa Major III is confirmed, standard thermal dark matter annihilation to $b\bar{b}$ final states would be ruled out for dark matter masses up to 4~TeV. The discovery of Ursa Major III, combined with recent tentative measurements of other high J-factor systems, suggests the exciting possibility that near-future data could produce transformative constraints on thermal dark matter.

\end{abstract}

\maketitle


\section{Introduction}
\label{sec:intro}
\vspace{-0.2cm}

Dwarf spheroidal galaxies (dSphs) have long been considered optimal targets in searches for dark matter annihilation due to their high dark matter densities and low backgrounds~\cite{Baltz:1999ra, Bergstrom:2000pn, Evans:2003sc, Strigari:2007at, 2011PhRvL.107x1303G}. Notably, observations indicate that the annihilation of Weakly Interactive Massive Particles (WIMPs) at the thermal relic cross-section~\cite{Steigman:2012nb} may produce detectable signals in $\gamma$-ray instruments such as the \Fermi\ Large Area Telescope (LAT)~\cite{Baltz:2008wd, Fermi-LAT:2010cni, Fermi-LAT:2013sme, Geringer-Sameth:2014qqa, Geringer-Sameth:2014yza, Fermi-LAT:2015att, Fermi-LAT:2016uux, Chiappo:2018mlt, Hoof:2018hyn, Linden:2019soa, Calore:2018sdx, Alvarez:2020cmw, McDaniel:2023bju, Li:2015kag, Li:2018kgy, Li:2021vqg}. 

Over the last decade, significant advancements in joint-likelihood analyses of the Milky Way dSph population have produced leading limits on dark matter annihilation. The most recent work by Ref.~\cite{McDaniel:2023bju}, ruled out thermal dark matter annihilation to the $b\bar{b}$ final state for dark matter masses below $\sim$80~GeV. Such analyses benefit from the fact that simultaneously analyzing many sources helps to mitigate uncertainties in the dark matter content or astrophysical background from any individual source. However, it remains possible that either the close proximity or fortuitously high dark matter content of a single dSph could single-handedly dominate dark matter limits. Indeed, such a scenario has previously been suggested for the Sagittarius~\cite{2010ApJ...725.1516L, Evans:2022zno} and Reticulum II dSphs~\cite{DES:2015txk, DES:2015tfc, Bonnivard:2015tta, Geringer-Sameth:2015lua, Hooper:2015ula}, as well as the Omega Cen globular cluster~\cite{Brown:2019whs, Reynoso-Cordova:2019biv, Evans:2021bsh}. 

\begin{figure}[]
\centering
\includegraphics[width = 0.48\textwidth]{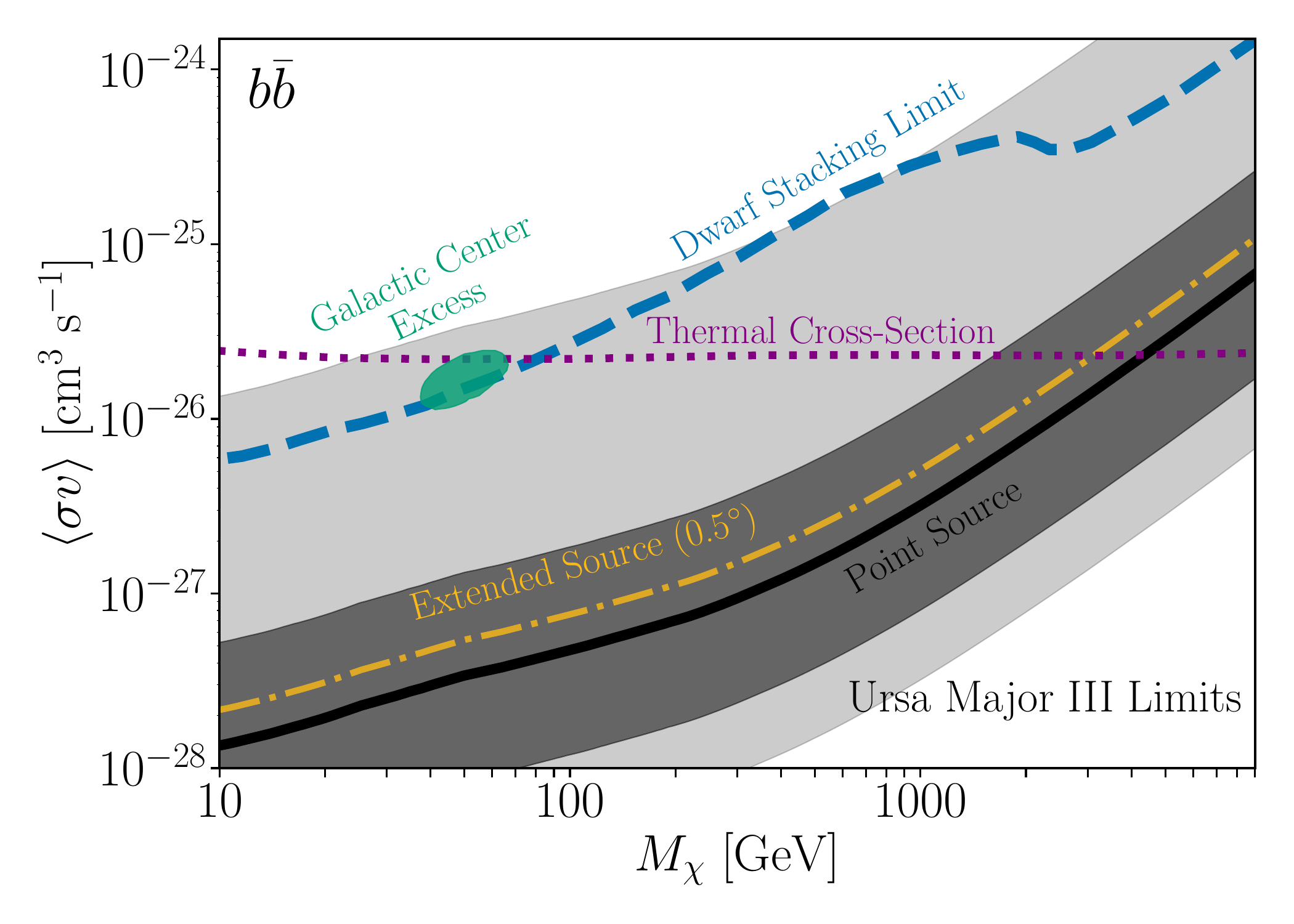}
\vspace{-0.5cm}
\caption{Upper limits on the annihilation cross-section from Ursa Major III (black solid), evaluated at the tentative J-factor of 10$^{21}$~GeV$^2$cm$^{-5}$~\cite{2023arXiv231110147S}, with both 0.6~dex error bands motivated by Ref.~\cite{2019MNRAS.482.3480P} (dark gray), and the 16--84\% error bands estimated by Ref.~\cite{2023arXiv231110134E} (light gray), assuming a point-source morphology. The yellow dash-dot line represents the upper limits of a 2D Gaussian distribution with a spatial width of 0.5$^{\circ}$, centered on the reported location of Ursa Major III. Considering Ursa Major III an extended source results in the limits $\sim$60\% weaker compared to the point-source consideration and comparable to the NFW profile consideration (limits not shown for plot clarity). We also show recent dSph joint-likelihood limits~\cite{McDaniel:2023bju} (dashed blue), a fit to the Galactic Center Excess~\cite{Calore:2014xka} (green region), and the thermal annihilation cross-section~\cite{Steigman:2012nb} (dotted purple). If the J-factor of Ursa Major III is confirmed, it would dominate indirect detection constraints, ruling out thermal WIMPs up to $\sim$2--4~TeV.}
\label{fig:comparison}
\end{figure}  

Recent studies have discovered a peculiar dSph that may fall into this vaunted class of sources~\cite{2023arXiv231110134E, 2023arXiv231110147S}. Ursa Major III/UNIONS I (hereafter, Ursa Major III), discovered by the Ultraviolet Near Infrared Optical Northern Survey, is already an extreme system, standing as the least luminous satellite ever discovered. Its dark matter content is unknown and depends on two separate arguments. The first is the measured stellar dispersion velocity of $3.7^{+1.4}_{-1.0}~{\rm km}~{\rm s}^{-1}$, which would require a large dark matter mass. However, this measurement has large uncertainties. Removing the largest velocity outlier from the dSph decreases the calculated dispersion velocity to $1.9^{+1.4}_{-1.1}$~km~s$^{-1}$, and removing an additional system eliminates it entirely. A second line of evidence, put forward by Ref.~\cite{2023arXiv231110134E}, indicates that Ursa Major III would be quickly disrupted by Milky Way tidal forces unless it were bound by a large dark gravitational component. The addition of a 10$^9$~M$_\odot$ dark matter halo would stabilize the system and induce a dispersion velocity of $\sim$1~km~s$^{-1}$.

The large mass and close proximity of Ursa Major III to Earth ($\sim$10$\pm$1~kpc) would suggest an unprecedented J-factor of $\sim$10$^{21}$~GeV$^2$~cm$^{-5}$. On top of these attributes, Ursa Major III lies at a high-galactic latitude ($l$=194.62$^\circ$, $b$=73.67$^\circ$) and far from any $\gamma$-ray sources (the nearest source is 4FGL J1135.1+3014, which is 1.16$^\circ$ away)~\cite{Fermi-LAT:2019yla}, making it an optimal target for \Fermi-LAT searches. 

In this \emph{paper}, we use 15 years of \Fermi-LAT data to search for a $\gamma$-ray excess consistent with Ursa Major III. Utilizing standard analysis cuts, we find no evidence of either point source or spatially extended emission consistent with the position of the dSph. Assuming a J-factor of 10$^{21}$~GeV$^2$~cm$^{-5}$, as motivated by Ref.~\cite{2023arXiv231110134E}, we set extremely strong limits on dark matter annihilation, which, if proven true, would rule out dark matter interpretations of the Galactic center excess~\cite{Daylan:2014rsa}, and more generally rule out standard thermal dark matter annihilation to $b\bar{b}$ final states for dark matter masses below 4~TeV (Fig.~\ref{fig:comparison}).

\section{Ursa Major III observations \& J-factor determination}
\label{sec:UM3-description}

The J-factors of dSphs depend on both their dark matter mass and density profile, as well as their distance from Earth~\cite{Gondolo:2004sc}. Because the dark matter parameters cannot be directly detected, stellar observations that trace the dark matter density, such as the stellar dispersion velocities and half-light radii, must be utilized to illuminate any underlying dark matter content~\cite{2007ApJ...670..313S}. The translation between these arguments can induce significant uncertainty, due to the fact that the stellar population is typically found only in the very inner portion of the dark matter halo. 

By far the most straightforward baryonic tracer of dSph J-factors is the distance, $D$, from the dSph to Earth. Because this parameter affects the dark matter flux quadratically, it has often been used as a simple predictor of the J-factor for dSphs with poorly constrained dark matter abundances. For such systems, models by Refs.~\cite{Fermi-LAT:2015att, Fermi-LAT:2016uux} utilized a relationship:

\begin{equation}
    {\rm log}_{10}\left(\frac{J_{{\rm pred}}}{J_0}\right) = -2~{\rm log}_{10}\left(\frac{D}{100~{\rm kpc}}\right)
\end{equation}

\noindent with a best-fit value of $J_0$~=~10$^{18.1}$~GeV$^2$~cm$^{-5}$~\cite{Fermi-LAT:2016uux}. In such a scenario, the close proximity of Ursa Major III would predict a J-factor of J$_{\rm pred}$~=~10$^{20.1}$~GeV$^2$~cm$^{-5}$. Based on spectroscopic measurements of better-determined dSphs, Ref.~\cite{Fermi-LAT:2016uux} tests intrinsic dispersions of 0.4, 0.6, and 0.8 dex for systems that only have J-factor estimations based on distance.

Recently, an in-depth analysis by Ref.~\cite{2019MNRAS.482.3480P} produced a more detailed model for the expected J-factor distribution from dSphs, arguing that the combined information from the line-of-sight velocity, $\sigma_{{\rm l.o.s.}}$, stellar half-light-radius, $r_{1/2}$, and distance can provide a much more accurate prediction of the dSph J-factor. Specifically, they find that the expected J-factor within 0.5$^\circ$ of the dSph center is best fit by:

\begin{equation}
{\rm \mathcal{J}}(0.5^\circ) = 10^{17.87} \left(\frac{\sigma_{{\rm l.o.s.}}}{5~{\rm km}~{\rm s}^{-1}}\right)^4\left(\frac{D}{100~{\rm kpc}}\right)^{-2}\left(\frac{r_{1/2}}{100~{\rm pc}}\right)^{-1}
\end{equation}

\noindent and provides a significantly improved fit to the J-factor estimation of dSphs, with a typical dispersion of only 0.1 dex. In this case, the primary uncertainty on the J-factor of Ursa Major III is the uncertainty in the line-of-sight dispersion velocity, which quartically enters the expected J-factor measurement. Using the statistical uncertainty of $\sigma_{{\rm l.o.s.}}$~=~3.7$^{+1.4}_{-1.0}$~km~s$^{-1}$, and taking this uncertainty in quadrature with the less important statistical uncertainties on the half-light radius (3$\pm$1~pc), distance (10$\pm$1~kpc), and the intrinsic dispersion, we roughly obtain a J-factor range of:

\begin{equation}
    {\rm log}_{10} \left(\frac{\mathcal{J}(0.5^\circ)}{{\rm GeV}^2~{\rm cm}^{-5}} \right) = 20.87^{+0.60}_{-0.58} 
    \label{eq:J}
\end{equation}

While this measurement is quantitatively motivated, it does not cover the true extent of J-factor uncertainties in Ursa Major III, due to the existence of significant systematic uncertainties stemming from uncertain stellar associations with this dSph. Notably, the dispersion velocity measurement of Ursa Major III is based on the identification of only 11 members and is strongly dependent on the association of two outlier member stars with the satellite~\cite{2023arXiv231110147S}. Removing the largest outlier star decreases the velocity dispersion to only $\sigma_{l.o.s.}$~=~1.9$^{+1.4}_{-1.1}$~km~s$^{-1}$, while also removing the second largest outlier eliminates the lower bound from the velocity dispersion measurement. Combining these errors within a statistical framework, Ref.~\cite{2023arXiv231110134E} finds a 16--84\% containment interval for the J-factor to be approximately 10$^{19}$ -- 10$^{22}$~GeV$^2$~cm$^{-5}$. 

\section{\Fermi-LAT data analysis}
\label{sec:lat}

The Large Area Telescope (LAT), onboard the \Fermi\ $\gamma$-ray observatory, is a pair-conversion telescope optimized for observing $\gamma$-rays with energies spanning $\sim$100~MeV to $\sim$500~GeV. Its field of view (FoV) covers 2.4~sr ($\sim$1/5 of the entire sky) and it can localize sources to an accuracy of a few arcminutes. A comprehensive description of the LAT specifications and its performance can be found in the dedicated instrument paper \cite{LATinstrument:2009}.

Here, we analyze 15 years of \Fermi-LAT data obtained from August 4, 2008 to August 4, 2023 and sourced from the publicly accessible \Fermi\ \textit{Science Support Center} (FSSC) website\footnote{\url{https://fermi.gsfc.nasa.gov/ssc/data/}, accessed on November 21, 2023.}. We utilize the open-source \texttt{fermipy} package (v.~1.2.0) \cite{fermipy:2017}, with an underlying dependence on \texttt{FermiTools} (v.~2.2.0)\footnote{\url{https://fermi.gsfc.nasa.gov/ssc/data/analysis/software/}, accessed on November 21, 2023.}. We consider an energy range of 500~MeV to 500~GeV spaced in 8 logarithmic bins per decade in energy. The observed $\gamma$-ray events are passed through the \texttt{P8R3\_SOURCE\_V3} event class selection. We make use of the standard models for the Galactic diffuse emission (\texttt{gll\_iem\_v07.fits}) and the isotropic component (\texttt{iso\_P8R3\_SOURCE\_V3\_v1.txt}) provided by the LAT Collaboration\footnote{\url{https://fermi.gsfc.nasa.gov/ssc/data/access/lat/BackgroundModels.html}, accessed on November 21, 2023.}. We consider a $10^{\circ} \times 10^{\circ}$ region of interest (RoI) centered at the position of \UMa\ (J2000: RA=174.71$^{\circ}$, Dec=31.08$^{\circ}$). To account for any leakage of $\gamma$-ray photons immediately outside the considered RoI, we model the background of point and extended sources using the 4FGL-DR3 catalog within a $15^{\circ} \times 15^{\circ}$ region centered at \UMa\ \cite{Fermi-LAT:2019yla, 4FGL-DR3:2022}. Any contamination from the Earth limb emission is accounted for by introducing the zenith angle cut of 100$^{\circ}$. 

In the standard \texttt{fermipy} analysis, the goodness of fit is quantified using the Test Statistic, $ TS = -2~{\rm log} \left(\mathcal{L}_0/\mathcal{L}_1\right)$, where $\mathcal{L}_0$ is the likelihood of the null hypothesis, and $\mathcal{L}_1$ is that of the alternative. Here, $\rm TS$ is a proxy for statistical significance and can be expressed in terms of the standard deviation, $\sigma$, as $\sqrt{\rm TS} \approx \sigma$ \cite{wilks1938}. However, studies of ``blank sky" locations with angular distributions similar to the dSph population have typically found that high-significance excesses are somewhat more common than predicted by Poisson fluctuations, a result which typically lowers the significance of some $\gamma$-ray excesses in the observed dSph population~\cite{Fermi-LAT:2013sme, Geringer-Sameth:2014qqa, Linden:2019soa}.

\begin{figure}[]
\centering
\includegraphics[width = 0.5\textwidth]{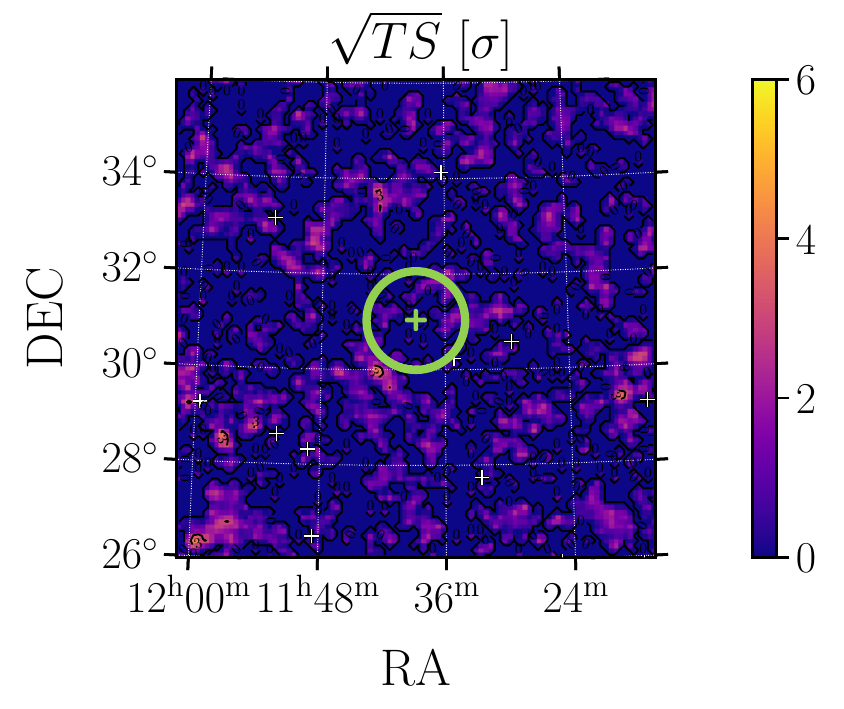}
\caption{Residual TS map of the 10$^\circ$$\times$10$^\circ$ RoI centered on \UMa\ (central green cross). Positions of sources removed in the data reduction process are shown as white crosses. The green circle shows 1$^{\circ}$ distance from the central point. We report no statistically significant excess in 15 years of \Fermi-LAT data in either the point-source or extended-source analysis at the position of \UMa.}
\label{fig:residual}
\end{figure}

\begin{figure}[]
\centering
\includegraphics[width = 0.5\textwidth]{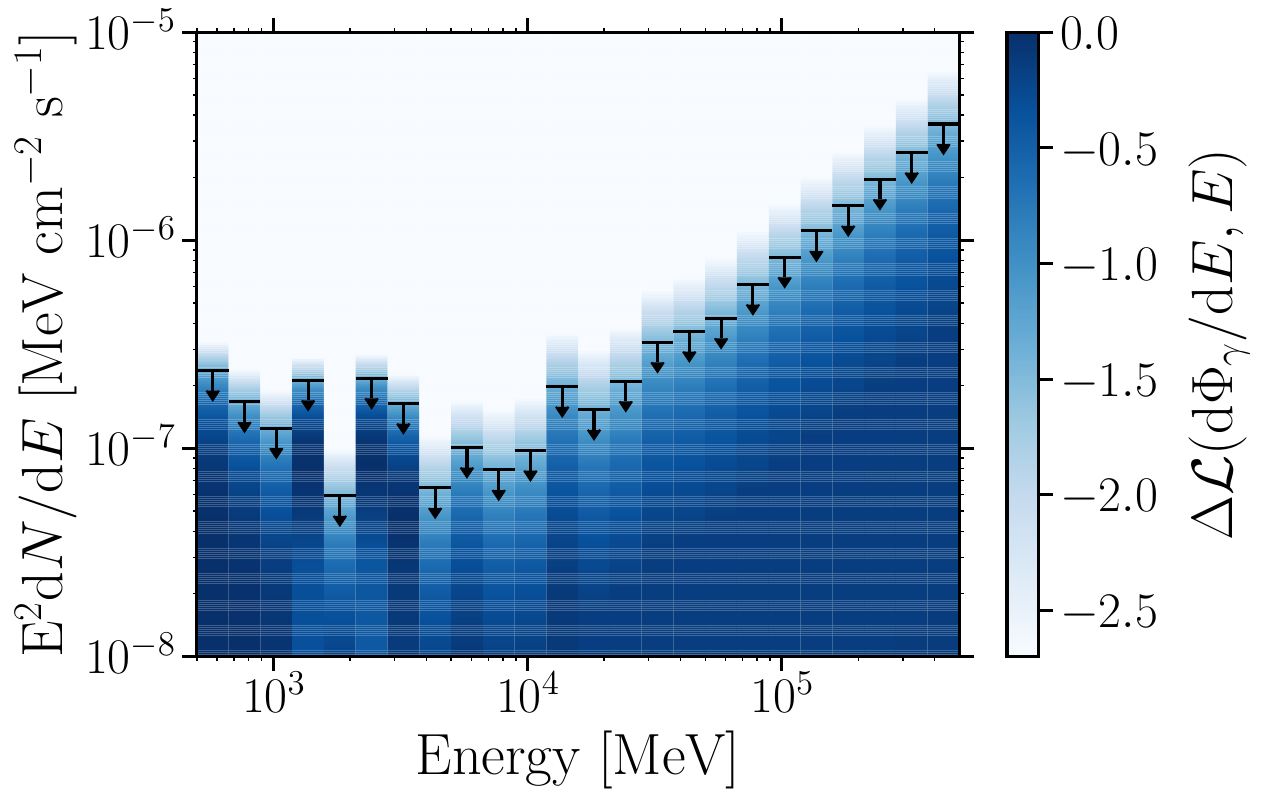}
\caption{The upper limit on the spectral energy distribution and the log-likelihood profile of \UMa\ between 500~MeV and 500~GeV under the assumption of a point source morphology. The arrows correspond to the flux upper limits at the 95-\% CL.}
\label{fig:flux}
\end{figure}  

Assuming a point-like emission from a dSph may (erroneously) produce more stringent limits on dark matter properties by a factor of $\sim2$--$2.5$ \cite{2022PhRvD.106l3032D}. While most dSphs are too faint to be spatially resolved by the LAT, considering their spatial extension becomes plausible---and crucial---for the nearby, large satellites. Numerous studies have considered various spatial morphologies to characterize the extended emission from dSphs in $\gamma$-rays (see, e.g., \cite{Fermi-LAT:2013sme, Fermi-LAT:2015att, Geringer-Sameth:2015lua}). Due to its proximity (heliocentric distance of $\sim10\pm 1$~kpc), \UMa~presents itself as a strong candidate for extended emission in the LAT. 

With that in mind, in addition to the point-source consideration, we also apply the \texttt{extension()} routine in \texttt{fermipy} to conduct a spatially resolved analysis centered at \UMa's reported position. This routine runs a likelihood ratio test that compares a point-like and an extended source model described by a chosen spatial morphology. The best-fit model for the extended fit is the one with the maximum likelihood value found by scanning through the likelihood profiles over the specified source width. 

First, we assume 2D Gaussian distributions with 0.1$^{\circ}$ and 0.5$^{\circ}$ spatial widths. The choice of 0.1$^{\circ}$ is motivated by the \UMa's half-light radius reported in \cite{2023arXiv231110147S}, which extends up to $\sim0.02^{\circ}$. This roughly translates into a $\sim0.1^{\circ}$ stellar tidal radius, which corresponds to---at least---the size of the dark matter tidal radius in a dSph system~\cite{2011MNRAS.411.2118A, Evans:2016xwx}. On the other hand, our choice of 0.5$^{\circ}$ stems from the half-light radius measurements of typical dSphs located at a distance similar to that of \UMa.  Indeed, the \UMa's half-light radius (3$\pm$1~pc) is relatively small compared to a typical dSph system, whose half-light radii can span a few tens to more than a hundred pcs (see, e.g., Table A1 in \cite{2019MNRAS.482.3480P}). 

In our spatially extended analysis, we also include a consideration of the Navarro-Frenk-White (NFW) density profile \cite{1996ApJ...462..563N}. Indeed, assuming the standard $\Lambda$CDM cosmology, a dark matter halo with a mass of 10$^{9}$~M$_{\odot}$ potentially seen in \UMa would result in a scale radius of $\sim 1.15$~kpc---corresponding to an angular size of $\sim$13$^{\circ}$. Of course, such a nearby, large system would experience high levels of tidal stripping, resulting in a disturbed dark matter distribution. Nevertheless, assuming the reported dynamical properties of \UMa\ (i.e., a circular velocity that peaks at $v_{\text{max}} = 23\,\text{km s}^{-1}$ at a radius $r_{\text{max}} = 3.0\,\text{kpc}$), followed by the conversion to the corresponding scale values (i.e., scale density $\rho_{\text{s}} = 2.6 \times 10^6\,\text{M}_{\odot}\,\text{kpc}^{-3}$ at a scale radius $r_{\text{s}} = 1.6\,\text{kpc}$) \cite{2008ApJ...686..262K}, and applying the \texttt{dmsky} module in \texttt{fermipy}, we produce a dark matter emission sky map extending up to $\sim$13$^{\circ}$ in the sky. Finally, we use the produced sky map to fit for dark matter emission in the $ 15^{\circ}\times15^{\circ}$ RoI centered at the reported \UMa\ position.
\begin{figure*}[t]
\centering
\includegraphics[width = 0.99\textwidth]{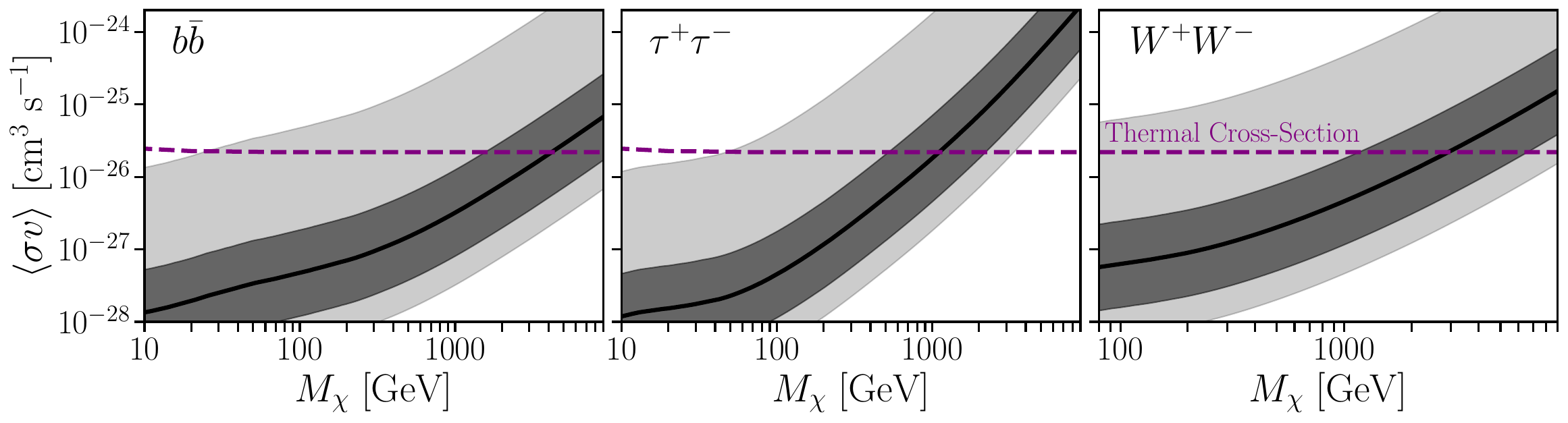}
\caption{Upper limits on the annihilation cross-section from Ursa Major III (black) as a function of dark matter mass, evaluated at the tentative J-factor value of 10$^{21}$~GeV$^2$cm$^{-5}$~\cite{2023arXiv231110147S} and assuming a point-source model for \UMa. We consider three annihilation channels: $b\bar{b}$ (left), $\tau^+\tau^-$ (center), and $W^+W^-$ (right). Also shown are 0.6~dex error bands motivated by Ref.~\cite{2019MNRAS.482.3480P} (dark gray) and the 16--84\% error bands estimated by Ref.~\cite{2023arXiv231110147S} (light gray). The canonical thermal annihilation cross-section is shown in purple~\cite{Steigman:2012nb}. The excluded WIMP masses reach up to $\sim$4~TeV, $\sim$1~TeV, and $\sim$2~TeV for $b\bar{b}$, $\tau^+\tau^-$, and $W^+W^-$ annihilation channels respectively.}
\label{fig:ul}
\end{figure*}

In each case, the flux of the source is calculated using the \texttt{sed()} method in \texttt{fermipy}, which conducts a fit in each energy bin assuming, in our case, a power-law distribution with the index of 2. The end result of the \texttt{sed()} method is a spectral energy distribution (SED) curve, $E^2 \dif N / \dif E$, and a delta log-likelihood profile, $\Delta \mathcal{L}\left(\dif \Phi_{\gamma}/\dif E, E\right)$.

Finally, once the flux is computed, we are set to conduct the conversion from the \UMa's SED into the dark matter space. The annihilation of dark matter produces a $\gamma$-ray flux, $\dif \Phi_{\gamma}/\dif E$, given by

\begin{equation}
\frac{\dif \Phi_{\gamma}}{\dif E}=\frac{1}{8\pi} \frac{\langle \sigma v \rangle}{M_{\chi}^2} \frac{\dif N}{\dif E} \times J,
\end{equation}
\noindent where $\langle \sigma v \rangle$ is the velocity-weighted annihilation cross-section, $M_{\chi}$ is the mass of the considered dark matter particle, $\dif N/\dif E$ is the number flux density per annihilation which depends on the considered dark matter mass and the annihilation channel (provided by \cite{Cirelli:2011}), and $J$ is the J-factor aforedescribed in Sec.~\ref{sec:UM3-description}. To constrain the dark matter parameter space, we compute the $TS$ values given by
\begin{equation}
    TS \left(\langle \sigma v \rangle, M_{\chi} \right) = -2~{\rm log} \frac{\mathcal{L}_0}{\mathcal{L}\left(\langle \sigma v \rangle, M_{\chi} \right)},
\end{equation}
\noindent where $\mathcal{L}_0$ is the likelihood of the null hypothesis (i.e., no dark matter present) and $\mathcal{L}\left(\langle \sigma v \rangle, M_{\chi} \right)$ is the likelihood of the model including dark matter. We consider a grid of 1000 logarithmically spaced dark matter masses within the 10~GeV to 10~TeV range and 100 logarithmically spaced cross-sections between $3\times10^{-29}$~cm$^3$s$^{-1}$ and $3\times10^{-24}$~cm$^3$s$^{-1}$. Under the assumption of a J-factor of 10$^{21}$~GeV$^2$~cm$^{-5}$, we test three different dark matter $\gamma$-ray spectra stemming from annihilation which primarily proceeds into $b\bar{b}$, $\tau^+\tau^-$, and $W^+W^-$ final states. Finally, by varying the $\langle \sigma v \rangle$ values, we determine the cross-section value for a specific dark matter mass for which the corresponding $TS$ value decreases by 2.71, resulting in the 95\% confidence-level (CL) limit on the dark matter annihilation cross-section.

\section{Results}
\label{sec:results}

In Fig.~\ref{fig:residual}, we show the residual (i.e., background-subtracted) TS map in our RoI, under the assumption of a   point-like source at the \UMa\ location. We report no statistically significant point-source emission in the 15 years of LAT data within the 500~MeV--500~GeV energy range at the position of \UMa. We also conduct extended-source modeling, assuming a 2D Gaussian distribution with a spatial extension of 0.1$^{\circ}$ and 0.5$^{\circ}$ at the \UMa\ position. Similarly, we report no statistically significant extended emission in this region in the \Fermi-LAT data for either of the input models, including an NFW profile.

Since no excess was found either in the point- or in the extended-source analysis, we proceed to calculate the flux upper limits assuming a point source behavior at the position of \UMa. The resulting 95\% CL flux upper limit is shown in Fig.~\ref{fig:flux}. We get consistent results for extensions of 0.1$^\circ$ and 0.5$^\circ$. The flux upper limits in the case of a 0.1$^\circ$ extension are negligibly higher, while for a 0.5$^\circ$ extension, which is larger than the value consistent with the half-light radius of Ursa Major III, we still get upper limits that are weaker only by about 60\%.

Finally, in Fig.~\ref{fig:ul} we report the upper limits on the dark matter mass and cross-sections for three different annihilation channels, $b\bar{b}$, $\tau^+\tau^-$, and $W^+W^-$. At the 95\% CL and assuming a J-factor value of 10$^{21}$~GeV$^2$~cm$^{-5}$, we exclude thermal annihilation cross-sections at dark matter masses up to $\sim$4~TeV, $\sim$1~TeV, and $\sim$2~TeV for dark matter annihilation into $b\bar{b}$, $\tau^+\tau^-$, and $W^+W^-$ respectively.

\section{Discussion \& Conclusions}
\label{sec:disc}

In this work, we have analyzed 15~years of \Fermi-LAT $\gamma$-ray data and set strong constraints on any $\gamma$-ray emission emanating from Ursa Major III. We have tested both point-source and extended-source templates for the $\gamma$-ray flux, which are compatible with interpretations of both partially stripped and unstripped dark matter profiles. Using a tentative J-factor estimate of 10$^{21}$~GeV$^2$~cm$^{-5}$, we show that the non-observation of Ursa Major III can set world-leading constraints on dark matter annihilation, potentially ruling out thermally annihilating WIMPs for dark matter masses up to 4~TeV assuming standard annihilation into $b\bar{b}$ final states.

We stress that the interpretation of these results depends critically on the J-factor (and indeed, existence) of dark matter within Ursa Major III. While the $\gamma$-ray flux constraints derived in this work are both robust and relatively simple (given the optimal location of Ursa Major III far from the galactic plane and any $\gamma$-ray point sources), the dark matter limits shown in this paper could entirely disappear if the large dark matter content of Ursa Major III is not confirmed in future experiments.

Thus, our result places Ursa Major III into an emerging category of systems (alongside Sagittarius, and Omega Cen), which may provide extremely strong constraints that rule out thermal WIMP annihilation --- if their tentative J-factor estimations are confirmed. Notably, only one of the above systems would be required to have a large dark matter component in order to strongly constrain the WIMP parameter space. Fortunately, upcoming survey instruments (e.g., Rubin Observatory~\cite{LSSTDarkMatterGroup:2019mwo, Ando:2019rvr}), and spectroscopic instruments (e.g., TMT, GMT, MSE, and ELT), are well-poised to provide detail population-level and spectroscopic verification of the dark matter content of dSphs over the next decade~\cite{2015RAA....15.1945S, 2016arXiv160600060M, 2018SPIE10700E..12F}, strengthening the potential for such surveys to strongly constrain or detect particle dark matter. 

In this sense, our results contribute to a possible transformation in the field of $\gamma$-ray indirect detection. The last decade has been marked primarily by incremental progress based on the increasing exposure of \FermiLAT and ACT observations along with improving models for the systematic uncertainties in $\gamma$-ray foreground measurements. However, the next decade of dark matter indirect detection may be dominated by optical observations and spectroscopic measurements of extreme dark matter targets that provide stringent dark matter limits without the need for detailed $\gamma$-ray modeling. Such sources renew the hope that near-future studies can definitively test thermal dark matter annihilation. 
\\

\acknowledgements

We thank Dan Hooper, Louis Strigari, and Alex Drlica-Wagner for helpful comments. MC and TL are supported by the Swedish Research Council under contract 2022-04283. TL also acknowledges support by the  European Research
Council under grant 742104 and the Swedish National Space
Agency under contract 117/19. 

This project made use of \texttt{Astropy} \cite{astropy:2022}, \texttt{Numpy} \citep{numpy:2020}, \texttt{Matplotlib} \cite{matplotlib:2007}, and \texttt{Pandas} \cite{pandas:2020} Python packages. The authors also acknowledge the use of public data from the \textit{Fermi Science Support Center} data archive.  


\bibliography{bibliograpy}

\providecommand{\noopsort}[1]{}\providecommand{\singleletter}[1]{#1}%
\begin{thebibliography}{59}%
\makeatletter
\providecommand \@ifxundefined [1]{%
 \@ifx{#1\undefined}
}%
\providecommand \@ifnum [1]{%
 \ifnum #1\expandafter \@firstoftwo
 \else \expandafter \@secondoftwo
 \fi
}%
\providecommand \@ifx [1]{%
 \ifx #1\expandafter \@firstoftwo
 \else \expandafter \@secondoftwo
 \fi
}%
\providecommand \natexlab [1]{#1}%
\providecommand \enquote  [1]{``#1''}%
\providecommand \bibnamefont  [1]{#1}%
\providecommand \bibfnamefont [1]{#1}%
\providecommand \citenamefont [1]{#1}%
\providecommand \href@noop [0]{\@secondoftwo}%
\providecommand \href [0]{\begingroup \@sanitize@url \@href}%
\providecommand \@href[1]{\@@startlink{#1}\@@href}%
\providecommand \@@href[1]{\endgroup#1\@@endlink}%
\providecommand \@sanitize@url [0]{\catcode `\\12\catcode `\$12\catcode `\&12\catcode `\#12\catcode `\^12\catcode `\_12\catcode `\%12\relax}%
\providecommand \@@startlink[1]{}%
\providecommand \@@endlink[0]{}%
\providecommand \url  [0]{\begingroup\@sanitize@url \@url }%
\providecommand \@url [1]{\endgroup\@href {#1}{\urlprefix }}%
\providecommand \urlprefix  [0]{URL }%
\providecommand \Eprint [0]{\href }%
\providecommand \doibase [0]{http://dx.doi.org/}%
\providecommand \selectlanguage [0]{\@gobble}%
\providecommand \bibinfo  [0]{\@secondoftwo}%
\providecommand \bibfield  [0]{\@secondoftwo}%
\providecommand \translation [1]{[#1]}%
\providecommand \BibitemOpen [0]{}%
\providecommand \bibitemStop [0]{}%
\providecommand \bibitemNoStop [0]{.\EOS\space}%
\providecommand \EOS [0]{\spacefactor3000\relax}%
\providecommand \BibitemShut  [1]{\csname bibitem#1\endcsname}%
\let\auto@bib@innerbib\@empty
\bibitem [{\citenamefont {Baltz}\ \emph {et~al.}(2000)\citenamefont {Baltz}, \citenamefont {Briot}, \citenamefont {Salati}, \citenamefont {Taillet},\ and\ \citenamefont {Silk}}]{Baltz:1999ra}%
  \BibitemOpen
  \bibfield  {author} {\bibinfo {author} {\bibfnamefont {E.~A.}\ \bibnamefont {Baltz}}, \bibinfo {author} {\bibfnamefont {C.}~\bibnamefont {Briot}}, \bibinfo {author} {\bibfnamefont {P.}~\bibnamefont {Salati}}, \bibinfo {author} {\bibfnamefont {R.}~\bibnamefont {Taillet}}, \ and\ \bibinfo {author} {\bibfnamefont {J.}~\bibnamefont {Silk}},\ }\href {\doibase 10.1103/PhysRevD.61.023514} {\bibfield  {journal} {\bibinfo  {journal} {Phys. Rev. D}\ }\textbf {\bibinfo {volume} {61}},\ \bibinfo {pages} {023514} (\bibinfo {year} {2000})},\ \Eprint {http://arxiv.org/abs/astro-ph/9909112} {arXiv:astro-ph/9909112} \BibitemShut {NoStop}%
\bibitem [{\citenamefont {Bergstr\"om}(2000)}]{Bergstrom:2000pn}%
  \BibitemOpen
  \bibfield  {author} {\bibinfo {author} {\bibfnamefont {L.}~\bibnamefont {Bergstr\"om}},\ }\href {\doibase 10.1088/0034-4885/63/5/2r3} {\bibfield  {journal} {\bibinfo  {journal} {Rept. Prog. Phys.}\ }\textbf {\bibinfo {volume} {63}},\ \bibinfo {pages} {793} (\bibinfo {year} {2000})},\ \Eprint {http://arxiv.org/abs/hep-ph/0002126} {arXiv:hep-ph/0002126} \BibitemShut {NoStop}%
\bibitem [{\citenamefont {Evans}\ \emph {et~al.}(2004)\citenamefont {Evans}, \citenamefont {Ferrer},\ and\ \citenamefont {Sarkar}}]{Evans:2003sc}%
  \BibitemOpen
  \bibfield  {author} {\bibinfo {author} {\bibfnamefont {N.~W.}\ \bibnamefont {Evans}}, \bibinfo {author} {\bibfnamefont {F.}~\bibnamefont {Ferrer}}, \ and\ \bibinfo {author} {\bibfnamefont {S.}~\bibnamefont {Sarkar}},\ }\href {\doibase 10.1103/PhysRevD.69.123501} {\bibfield  {journal} {\bibinfo  {journal} {Phys. Rev. D}\ }\textbf {\bibinfo {volume} {69}},\ \bibinfo {pages} {123501} (\bibinfo {year} {2004})},\ \Eprint {http://arxiv.org/abs/astro-ph/0311145} {arXiv:astro-ph/0311145} \BibitemShut {NoStop}%
\bibitem [{\citenamefont {Strigari}\ \emph {et~al.}(2008)\citenamefont {Strigari}, \citenamefont {Koushiappas}, \citenamefont {Bullock}, \citenamefont {Kaplinghat}, \citenamefont {Simon}, \citenamefont {Geha},\ and\ \citenamefont {Willman}}]{Strigari:2007at}%
  \BibitemOpen
  \bibfield  {author} {\bibinfo {author} {\bibfnamefont {L.~E.}\ \bibnamefont {Strigari}}, \bibinfo {author} {\bibfnamefont {S.~M.}\ \bibnamefont {Koushiappas}}, \bibinfo {author} {\bibfnamefont {J.~S.}\ \bibnamefont {Bullock}}, \bibinfo {author} {\bibfnamefont {M.}~\bibnamefont {Kaplinghat}}, \bibinfo {author} {\bibfnamefont {J.~D.}\ \bibnamefont {Simon}}, \bibinfo {author} {\bibfnamefont {M.}~\bibnamefont {Geha}}, \ and\ \bibinfo {author} {\bibfnamefont {B.}~\bibnamefont {Willman}},\ }\href {\doibase 10.1086/529488} {\bibfield  {journal} {\bibinfo  {journal} {Astrophys. J.}\ }\textbf {\bibinfo {volume} {678}},\ \bibinfo {pages} {614} (\bibinfo {year} {2008})},\ \Eprint {http://arxiv.org/abs/0709.1510} {arXiv:0709.1510 [astro-ph]} \BibitemShut {NoStop}%
\bibitem [{\citenamefont {{Geringer-Sameth}}\ and\ \citenamefont {{Koushiappas}}(2011)}]{2011PhRvL.107x1303G}%
  \BibitemOpen
  \bibfield  {author} {\bibinfo {author} {\bibfnamefont {A.}~\bibnamefont {{Geringer-Sameth}}}\ and\ \bibinfo {author} {\bibfnamefont {S.~M.}\ \bibnamefont {{Koushiappas}}},\ }\href {\doibase 10.1103/PhysRevLett.107.241303} {\bibfield  {journal} {\bibinfo  {journal} {\prl}\ }\textbf {\bibinfo {volume} {107}},\ \bibinfo {eid} {241303} (\bibinfo {year} {2011})},\ \Eprint {http://arxiv.org/abs/1108.2914} {arXiv:1108.2914 [astro-ph.CO]} \BibitemShut {NoStop}%
\bibitem [{\citenamefont {Steigman}\ \emph {et~al.}(2012)\citenamefont {Steigman}, \citenamefont {Dasgupta},\ and\ \citenamefont {Beacom}}]{Steigman:2012nb}%
  \BibitemOpen
  \bibfield  {author} {\bibinfo {author} {\bibfnamefont {G.}~\bibnamefont {Steigman}}, \bibinfo {author} {\bibfnamefont {B.}~\bibnamefont {Dasgupta}}, \ and\ \bibinfo {author} {\bibfnamefont {J.~F.}\ \bibnamefont {Beacom}},\ }\href {\doibase 10.1103/PhysRevD.86.023506} {\bibfield  {journal} {\bibinfo  {journal} {Phys. Rev. D}\ }\textbf {\bibinfo {volume} {86}},\ \bibinfo {pages} {023506} (\bibinfo {year} {2012})},\ \Eprint {http://arxiv.org/abs/1204.3622} {arXiv:1204.3622 [hep-ph]} \BibitemShut {NoStop}%
\bibitem [{\citenamefont {Baltz}\ \emph {et~al.}(2008)\citenamefont {Baltz} \emph {et~al.}}]{Baltz:2008wd}%
  \BibitemOpen
  \bibfield  {author} {\bibinfo {author} {\bibfnamefont {E.~A.}\ \bibnamefont {Baltz}} \emph {et~al.},\ }\href {\doibase 10.1088/1475-7516/2008/07/013} {\bibfield  {journal} {\bibinfo  {journal} {JCAP}\ }\textbf {\bibinfo {volume} {07}},\ \bibinfo {pages} {013} (\bibinfo {year} {2008})},\ \Eprint {http://arxiv.org/abs/0806.2911} {arXiv:0806.2911 [astro-ph]} \BibitemShut {NoStop}%
\bibitem [{\citenamefont {Abdo}\ \emph {et~al.}(2010)\citenamefont {Abdo} \emph {et~al.}}]{Fermi-LAT:2010cni}%
  \BibitemOpen
  \bibfield  {author} {\bibinfo {author} {\bibfnamefont {A.~A.}\ \bibnamefont {Abdo}} \emph {et~al.} (\bibinfo {collaboration} {Fermi-LAT}),\ }\href {\doibase 10.1088/0004-637X/712/1/147} {\bibfield  {journal} {\bibinfo  {journal} {Astrophys. J.}\ }\textbf {\bibinfo {volume} {712}},\ \bibinfo {pages} {147} (\bibinfo {year} {2010})},\ \Eprint {http://arxiv.org/abs/1001.4531} {arXiv:1001.4531 [astro-ph.CO]} \BibitemShut {NoStop}%
\bibitem [{\citenamefont {Ackermann}\ \emph {et~al.}(2014)\citenamefont {Ackermann} \emph {et~al.}}]{Fermi-LAT:2013sme}%
  \BibitemOpen
  \bibfield  {author} {\bibinfo {author} {\bibfnamefont {M.}~\bibnamefont {Ackermann}} \emph {et~al.} (\bibinfo {collaboration} {Fermi-LAT}),\ }\href {\doibase 10.1103/PhysRevD.89.042001} {\bibfield  {journal} {\bibinfo  {journal} {Phys. Rev. D}\ }\textbf {\bibinfo {volume} {89}},\ \bibinfo {pages} {042001} (\bibinfo {year} {2014})},\ \Eprint {http://arxiv.org/abs/1310.0828} {arXiv:1310.0828 [astro-ph.HE]} \BibitemShut {NoStop}%
\bibitem [{\citenamefont {Geringer-Sameth}\ \emph {et~al.}(2015{\natexlab{a}})\citenamefont {Geringer-Sameth}, \citenamefont {Koushiappas},\ and\ \citenamefont {Walker}}]{Geringer-Sameth:2014qqa}%
  \BibitemOpen
  \bibfield  {author} {\bibinfo {author} {\bibfnamefont {A.}~\bibnamefont {Geringer-Sameth}}, \bibinfo {author} {\bibfnamefont {S.~M.}\ \bibnamefont {Koushiappas}}, \ and\ \bibinfo {author} {\bibfnamefont {M.~G.}\ \bibnamefont {Walker}},\ }\href {\doibase 10.1103/PhysRevD.91.083535} {\bibfield  {journal} {\bibinfo  {journal} {Phys. Rev. D}\ }\textbf {\bibinfo {volume} {91}},\ \bibinfo {pages} {083535} (\bibinfo {year} {2015}{\natexlab{a}})},\ \Eprint {http://arxiv.org/abs/1410.2242} {arXiv:1410.2242 [astro-ph.CO]} \BibitemShut {NoStop}%
\bibitem [{\citenamefont {Geringer-Sameth}\ \emph {et~al.}(2015{\natexlab{b}})\citenamefont {Geringer-Sameth}, \citenamefont {Koushiappas},\ and\ \citenamefont {Walker}}]{Geringer-Sameth:2014yza}%
  \BibitemOpen
  \bibfield  {author} {\bibinfo {author} {\bibfnamefont {A.}~\bibnamefont {Geringer-Sameth}}, \bibinfo {author} {\bibfnamefont {S.~M.}\ \bibnamefont {Koushiappas}}, \ and\ \bibinfo {author} {\bibfnamefont {M.}~\bibnamefont {Walker}},\ }\href {\doibase 10.1088/0004-637X/801/2/74} {\bibfield  {journal} {\bibinfo  {journal} {Astrophys. J.}\ }\textbf {\bibinfo {volume} {801}},\ \bibinfo {pages} {74} (\bibinfo {year} {2015}{\natexlab{b}})},\ \Eprint {http://arxiv.org/abs/1408.0002} {arXiv:1408.0002 [astro-ph.CO]} \BibitemShut {NoStop}%
\bibitem [{\citenamefont {Ackermann}\ \emph {et~al.}(2015)\citenamefont {Ackermann} \emph {et~al.}}]{Fermi-LAT:2015att}%
  \BibitemOpen
  \bibfield  {author} {\bibinfo {author} {\bibfnamefont {M.}~\bibnamefont {Ackermann}} \emph {et~al.} (\bibinfo {collaboration} {Fermi-LAT}),\ }\href {\doibase 10.1103/PhysRevLett.115.231301} {\bibfield  {journal} {\bibinfo  {journal} {Phys. Rev. Lett.}\ }\textbf {\bibinfo {volume} {115}},\ \bibinfo {pages} {231301} (\bibinfo {year} {2015})},\ \Eprint {http://arxiv.org/abs/1503.02641} {arXiv:1503.02641 [astro-ph.HE]} \BibitemShut {NoStop}%
\bibitem [{\citenamefont {Albert}\ \emph {et~al.}(2017)\citenamefont {Albert} \emph {et~al.}}]{Fermi-LAT:2016uux}%
  \BibitemOpen
  \bibfield  {author} {\bibinfo {author} {\bibfnamefont {A.}~\bibnamefont {Albert}} \emph {et~al.} (\bibinfo {collaboration} {Fermi-LAT, DES}),\ }\href {\doibase 10.3847/1538-4357/834/2/110} {\bibfield  {journal} {\bibinfo  {journal} {Astrophys. J.}\ }\textbf {\bibinfo {volume} {834}},\ \bibinfo {pages} {110} (\bibinfo {year} {2017})},\ \Eprint {http://arxiv.org/abs/1611.03184} {arXiv:1611.03184 [astro-ph.HE]} \BibitemShut {NoStop}%
\bibitem [{\citenamefont {Chiappo}\ \emph {et~al.}(2019)\citenamefont {Chiappo}, \citenamefont {Cohen-Tanugi}, \citenamefont {Conrad},\ and\ \citenamefont {Strigari}}]{Chiappo:2018mlt}%
  \BibitemOpen
  \bibfield  {author} {\bibinfo {author} {\bibfnamefont {A.}~\bibnamefont {Chiappo}}, \bibinfo {author} {\bibfnamefont {J.}~\bibnamefont {Cohen-Tanugi}}, \bibinfo {author} {\bibfnamefont {J.}~\bibnamefont {Conrad}}, \ and\ \bibinfo {author} {\bibfnamefont {L.~E.}\ \bibnamefont {Strigari}},\ }\href {\doibase 10.1093/mnras/stz1871} {\bibfield  {journal} {\bibinfo  {journal} {Mon. Not. Roy. Astron. Soc.}\ }\textbf {\bibinfo {volume} {488}},\ \bibinfo {pages} {2616} (\bibinfo {year} {2019})},\ \Eprint {http://arxiv.org/abs/1810.09917} {arXiv:1810.09917 [astro-ph.GA]} \BibitemShut {NoStop}%
\bibitem [{\citenamefont {Hoof}\ \emph {et~al.}(2020)\citenamefont {Hoof}, \citenamefont {Geringer-Sameth},\ and\ \citenamefont {Trotta}}]{Hoof:2018hyn}%
  \BibitemOpen
  \bibfield  {author} {\bibinfo {author} {\bibfnamefont {S.}~\bibnamefont {Hoof}}, \bibinfo {author} {\bibfnamefont {A.}~\bibnamefont {Geringer-Sameth}}, \ and\ \bibinfo {author} {\bibfnamefont {R.}~\bibnamefont {Trotta}},\ }\href {\doibase 10.1088/1475-7516/2020/02/012} {\bibfield  {journal} {\bibinfo  {journal} {JCAP}\ }\textbf {\bibinfo {volume} {02}},\ \bibinfo {pages} {012} (\bibinfo {year} {2020})},\ \Eprint {http://arxiv.org/abs/1812.06986} {arXiv:1812.06986 [astro-ph.CO]} \BibitemShut {NoStop}%
\bibitem [{\citenamefont {Linden}(2020)}]{Linden:2019soa}%
  \BibitemOpen
  \bibfield  {author} {\bibinfo {author} {\bibfnamefont {T.}~\bibnamefont {Linden}},\ }\href {\doibase 10.1103/PhysRevD.101.043017} {\bibfield  {journal} {\bibinfo  {journal} {Phys. Rev. D}\ }\textbf {\bibinfo {volume} {101}},\ \bibinfo {pages} {043017} (\bibinfo {year} {2020})},\ \Eprint {http://arxiv.org/abs/1905.11992} {arXiv:1905.11992 [astro-ph.HE]} \BibitemShut {NoStop}%
\bibitem [{\citenamefont {Calore}\ \emph {et~al.}(2018)\citenamefont {Calore}, \citenamefont {Serpico},\ and\ \citenamefont {Zaldivar}}]{Calore:2018sdx}%
  \BibitemOpen
  \bibfield  {author} {\bibinfo {author} {\bibfnamefont {F.}~\bibnamefont {Calore}}, \bibinfo {author} {\bibfnamefont {P.~D.}\ \bibnamefont {Serpico}}, \ and\ \bibinfo {author} {\bibfnamefont {B.}~\bibnamefont {Zaldivar}},\ }\href {\doibase 10.1088/1475-7516/2018/10/029} {\bibfield  {journal} {\bibinfo  {journal} {JCAP}\ }\textbf {\bibinfo {volume} {10}},\ \bibinfo {pages} {029} (\bibinfo {year} {2018})},\ \Eprint {http://arxiv.org/abs/1803.05508} {arXiv:1803.05508 [astro-ph.HE]} \BibitemShut {NoStop}%
\bibitem [{\citenamefont {Alvarez}\ \emph {et~al.}(2020)\citenamefont {Alvarez}, \citenamefont {Calore}, \citenamefont {Genina}, \citenamefont {Read}, \citenamefont {Serpico},\ and\ \citenamefont {Zaldivar}}]{Alvarez:2020cmw}%
  \BibitemOpen
  \bibfield  {author} {\bibinfo {author} {\bibfnamefont {A.}~\bibnamefont {Alvarez}}, \bibinfo {author} {\bibfnamefont {F.}~\bibnamefont {Calore}}, \bibinfo {author} {\bibfnamefont {A.}~\bibnamefont {Genina}}, \bibinfo {author} {\bibfnamefont {J.}~\bibnamefont {Read}}, \bibinfo {author} {\bibfnamefont {P.~D.}\ \bibnamefont {Serpico}}, \ and\ \bibinfo {author} {\bibfnamefont {B.}~\bibnamefont {Zaldivar}},\ }\href {\doibase 10.1088/1475-7516/2020/09/004} {\bibfield  {journal} {\bibinfo  {journal} {JCAP}\ }\textbf {\bibinfo {volume} {09}},\ \bibinfo {pages} {004} (\bibinfo {year} {2020})},\ \Eprint {http://arxiv.org/abs/2002.01229} {arXiv:2002.01229 [astro-ph.HE]} \BibitemShut {NoStop}%
\bibitem [{\citenamefont {McDaniel}\ \emph {et~al.}(2023)\citenamefont {McDaniel}, \citenamefont {Ajello}, \citenamefont {Karwin}, \citenamefont {Di~Mauro}, \citenamefont {Drlica-Wagner},\ and\ \citenamefont {S\`anchez-Conde}}]{McDaniel:2023bju}%
  \BibitemOpen
  \bibfield  {author} {\bibinfo {author} {\bibfnamefont {A.}~\bibnamefont {McDaniel}}, \bibinfo {author} {\bibfnamefont {M.}~\bibnamefont {Ajello}}, \bibinfo {author} {\bibfnamefont {C.~M.}\ \bibnamefont {Karwin}}, \bibinfo {author} {\bibfnamefont {M.}~\bibnamefont {Di~Mauro}}, \bibinfo {author} {\bibfnamefont {A.}~\bibnamefont {Drlica-Wagner}}, \ and\ \bibinfo {author} {\bibfnamefont {M.}~\bibnamefont {S\`anchez-Conde}},\ }\href@noop {} {\  (\bibinfo {year} {2023})},\ \Eprint {http://arxiv.org/abs/2311.04982} {arXiv:2311.04982 [astro-ph.HE]} \BibitemShut {NoStop}%
\bibitem [{\citenamefont {Li}\ \emph {et~al.}(2016)\citenamefont {Li}, \citenamefont {Liang}, \citenamefont {Duan}, \citenamefont {Shen}, \citenamefont {Huang}, \citenamefont {Li}, \citenamefont {Fan}, \citenamefont {Liao}, \citenamefont {Feng},\ and\ \citenamefont {Chang}}]{Li:2015kag}%
  \BibitemOpen
  \bibfield  {author} {\bibinfo {author} {\bibfnamefont {S.}~\bibnamefont {Li}}, \bibinfo {author} {\bibfnamefont {Y.-F.}\ \bibnamefont {Liang}}, \bibinfo {author} {\bibfnamefont {K.-K.}\ \bibnamefont {Duan}}, \bibinfo {author} {\bibfnamefont {Z.-Q.}\ \bibnamefont {Shen}}, \bibinfo {author} {\bibfnamefont {X.}~\bibnamefont {Huang}}, \bibinfo {author} {\bibfnamefont {X.}~\bibnamefont {Li}}, \bibinfo {author} {\bibfnamefont {Y.-Z.}\ \bibnamefont {Fan}}, \bibinfo {author} {\bibfnamefont {N.-H.}\ \bibnamefont {Liao}}, \bibinfo {author} {\bibfnamefont {L.}~\bibnamefont {Feng}}, \ and\ \bibinfo {author} {\bibfnamefont {J.}~\bibnamefont {Chang}},\ }\href {\doibase 10.1103/PhysRevD.93.043518} {\bibfield  {journal} {\bibinfo  {journal} {Phys. Rev. D}\ }\textbf {\bibinfo {volume} {93}},\ \bibinfo {pages} {043518} (\bibinfo {year} {2016})},\ \Eprint {http://arxiv.org/abs/1511.09252} {arXiv:1511.09252 [astro-ph.HE]} \BibitemShut {NoStop}%
\bibitem [{\citenamefont {Li}\ \emph {et~al.}(2018)\citenamefont {Li} \emph {et~al.}}]{Li:2018kgy}%
  \BibitemOpen
  \bibfield  {author} {\bibinfo {author} {\bibfnamefont {S.}~\bibnamefont {Li}} \emph {et~al.},\ }\href {\doibase 10.1103/PhysRevD.97.122001} {\bibfield  {journal} {\bibinfo  {journal} {Phys. Rev. D}\ }\textbf {\bibinfo {volume} {97}},\ \bibinfo {pages} {122001} (\bibinfo {year} {2018})},\ \Eprint {http://arxiv.org/abs/1805.06612} {arXiv:1805.06612 [astro-ph.HE]} \BibitemShut {NoStop}%
\bibitem [{\citenamefont {Li}\ \emph {et~al.}(2021)\citenamefont {Li}, \citenamefont {Liang},\ and\ \citenamefont {Fan}}]{Li:2021vqg}%
  \BibitemOpen
  \bibfield  {author} {\bibinfo {author} {\bibfnamefont {S.}~\bibnamefont {Li}}, \bibinfo {author} {\bibfnamefont {Y.-F.}\ \bibnamefont {Liang}}, \ and\ \bibinfo {author} {\bibfnamefont {Y.-Z.}\ \bibnamefont {Fan}},\ }\href {\doibase 10.1103/PhysRevD.104.083037} {\bibfield  {journal} {\bibinfo  {journal} {Phys. Rev. D}\ }\textbf {\bibinfo {volume} {104}},\ \bibinfo {pages} {083037} (\bibinfo {year} {2021})},\ \Eprint {http://arxiv.org/abs/2110.01157} {arXiv:2110.01157 [astro-ph.HE]} \BibitemShut {NoStop}%
\bibitem [{\citenamefont {{{\L}okas}}\ \emph {et~al.}(2010)\citenamefont {{{\L}okas}}, \citenamefont {{Kazantzidis}}, \citenamefont {{Majewski}}, \citenamefont {{Law}}, \citenamefont {{Mayer}},\ and\ \citenamefont {{Frinchaboy}}}]{2010ApJ...725.1516L}%
  \BibitemOpen
  \bibfield  {author} {\bibinfo {author} {\bibfnamefont {E.~L.}\ \bibnamefont {{{\L}okas}}}, \bibinfo {author} {\bibfnamefont {S.}~\bibnamefont {{Kazantzidis}}}, \bibinfo {author} {\bibfnamefont {S.~R.}\ \bibnamefont {{Majewski}}}, \bibinfo {author} {\bibfnamefont {D.~R.}\ \bibnamefont {{Law}}}, \bibinfo {author} {\bibfnamefont {L.}~\bibnamefont {{Mayer}}}, \ and\ \bibinfo {author} {\bibfnamefont {P.~M.}\ \bibnamefont {{Frinchaboy}}},\ }\href {\doibase 10.1088/0004-637X/725/2/1516} {\bibfield  {journal} {\bibinfo  {journal} {\apj}\ }\textbf {\bibinfo {volume} {725}},\ \bibinfo {pages} {1516} (\bibinfo {year} {2010})},\ \Eprint {http://arxiv.org/abs/1008.3464} {arXiv:1008.3464 [astro-ph.CO]} \BibitemShut {NoStop}%
\bibitem [{\citenamefont {Evans}\ \emph {et~al.}(2023)\citenamefont {Evans}, \citenamefont {Strigari}, \citenamefont {Svenborn}, \citenamefont {Albert}, \citenamefont {Harding}, \citenamefont {Hooper}, \citenamefont {Linden},\ and\ \citenamefont {Pace}}]{Evans:2022zno}%
  \BibitemOpen
  \bibfield  {author} {\bibinfo {author} {\bibfnamefont {A.~J.}\ \bibnamefont {Evans}}, \bibinfo {author} {\bibfnamefont {L.~E.}\ \bibnamefont {Strigari}}, \bibinfo {author} {\bibfnamefont {O.}~\bibnamefont {Svenborn}}, \bibinfo {author} {\bibfnamefont {A.}~\bibnamefont {Albert}}, \bibinfo {author} {\bibfnamefont {J.~P.}\ \bibnamefont {Harding}}, \bibinfo {author} {\bibfnamefont {D.}~\bibnamefont {Hooper}}, \bibinfo {author} {\bibfnamefont {T.}~\bibnamefont {Linden}}, \ and\ \bibinfo {author} {\bibfnamefont {A.~B.}\ \bibnamefont {Pace}},\ }\href {\doibase 10.1093/mnras/stad2074} {\bibfield  {journal} {\bibinfo  {journal} {Mon. Not. Roy. Astron. Soc.}\ }\textbf {\bibinfo {volume} {524}},\ \bibinfo {pages} {4574} (\bibinfo {year} {2023})},\ \Eprint {http://arxiv.org/abs/2212.08194} {arXiv:2212.08194 [astro-ph.HE]} \BibitemShut {NoStop}%
\bibitem [{\citenamefont {Bechtol}\ \emph {et~al.}(2015)\citenamefont {Bechtol} \emph {et~al.}}]{DES:2015txk}%
  \BibitemOpen
  \bibfield  {author} {\bibinfo {author} {\bibfnamefont {K.}~\bibnamefont {Bechtol}} \emph {et~al.} (\bibinfo {collaboration} {DES}),\ }\href {\doibase 10.1088/0004-637X/807/1/50} {\bibfield  {journal} {\bibinfo  {journal} {Astrophys. J.}\ }\textbf {\bibinfo {volume} {807}},\ \bibinfo {pages} {50} (\bibinfo {year} {2015})},\ \Eprint {http://arxiv.org/abs/1503.02584} {arXiv:1503.02584 [astro-ph.GA]} \BibitemShut {NoStop}%
\bibitem [{\citenamefont {Simon}\ \emph {et~al.}(2015)\citenamefont {Simon} \emph {et~al.}}]{DES:2015tfc}%
  \BibitemOpen
  \bibfield  {author} {\bibinfo {author} {\bibfnamefont {J.~D.}\ \bibnamefont {Simon}} \emph {et~al.} (\bibinfo {collaboration} {DES}),\ }\href {\doibase 10.1088/0004-637X/808/1/95} {\bibfield  {journal} {\bibinfo  {journal} {Astrophys. J.}\ }\textbf {\bibinfo {volume} {808}},\ \bibinfo {pages} {95} (\bibinfo {year} {2015})},\ \Eprint {http://arxiv.org/abs/1504.02889} {arXiv:1504.02889 [astro-ph.GA]} \BibitemShut {NoStop}%
\bibitem [{\citenamefont {Bonnivard}\ \emph {et~al.}(2015)\citenamefont {Bonnivard}, \citenamefont {Combet}, \citenamefont {Maurin}, \citenamefont {Geringer-Sameth}, \citenamefont {Koushiappas}, \citenamefont {Walker}, \citenamefont {Mateo}, \citenamefont {Olszewski},\ and\ \citenamefont {Bailey~III}}]{Bonnivard:2015tta}%
  \BibitemOpen
  \bibfield  {author} {\bibinfo {author} {\bibfnamefont {V.}~\bibnamefont {Bonnivard}}, \bibinfo {author} {\bibfnamefont {C.}~\bibnamefont {Combet}}, \bibinfo {author} {\bibfnamefont {D.}~\bibnamefont {Maurin}}, \bibinfo {author} {\bibfnamefont {A.}~\bibnamefont {Geringer-Sameth}}, \bibinfo {author} {\bibfnamefont {S.~M.}\ \bibnamefont {Koushiappas}}, \bibinfo {author} {\bibfnamefont {M.~G.}\ \bibnamefont {Walker}}, \bibinfo {author} {\bibfnamefont {M.}~\bibnamefont {Mateo}}, \bibinfo {author} {\bibfnamefont {E.~W.}\ \bibnamefont {Olszewski}}, \ and\ \bibinfo {author} {\bibfnamefont {J.~I.}\ \bibnamefont {Bailey~III}},\ }\href {\doibase 10.1088/2041-8205/808/2/L36} {\bibfield  {journal} {\bibinfo  {journal} {Astrophys. J. Lett.}\ }\textbf {\bibinfo {volume} {808}},\ \bibinfo {pages} {L36} (\bibinfo {year} {2015})},\ \Eprint {http://arxiv.org/abs/1504.03309} {arXiv:1504.03309 [astro-ph.HE]} \BibitemShut {NoStop}%
\bibitem [{\citenamefont {Geringer-Sameth}\ \emph {et~al.}(2015{\natexlab{c}})\citenamefont {Geringer-Sameth}, \citenamefont {Walker}, \citenamefont {Koushiappas}, \citenamefont {Koposov}, \citenamefont {Belokurov}, \citenamefont {Torrealba},\ and\ \citenamefont {Evans}}]{Geringer-Sameth:2015lua}%
  \BibitemOpen
  \bibfield  {author} {\bibinfo {author} {\bibfnamefont {A.}~\bibnamefont {Geringer-Sameth}}, \bibinfo {author} {\bibfnamefont {M.~G.}\ \bibnamefont {Walker}}, \bibinfo {author} {\bibfnamefont {S.~M.}\ \bibnamefont {Koushiappas}}, \bibinfo {author} {\bibfnamefont {S.~E.}\ \bibnamefont {Koposov}}, \bibinfo {author} {\bibfnamefont {V.}~\bibnamefont {Belokurov}}, \bibinfo {author} {\bibfnamefont {G.}~\bibnamefont {Torrealba}}, \ and\ \bibinfo {author} {\bibfnamefont {N.~W.}\ \bibnamefont {Evans}},\ }\href {\doibase 10.1103/PhysRevLett.115.081101} {\bibfield  {journal} {\bibinfo  {journal} {Phys. Rev. Lett.}\ }\textbf {\bibinfo {volume} {115}},\ \bibinfo {pages} {081101} (\bibinfo {year} {2015}{\natexlab{c}})},\ \Eprint {http://arxiv.org/abs/1503.02320} {arXiv:1503.02320 [astro-ph.HE]} \BibitemShut {NoStop}%
\bibitem [{\citenamefont {Hooper}\ and\ \citenamefont {Linden}(2015)}]{Hooper:2015ula}%
  \BibitemOpen
  \bibfield  {author} {\bibinfo {author} {\bibfnamefont {D.}~\bibnamefont {Hooper}}\ and\ \bibinfo {author} {\bibfnamefont {T.}~\bibnamefont {Linden}},\ }\href {\doibase 10.1088/1475-7516/2015/09/016} {\bibfield  {journal} {\bibinfo  {journal} {JCAP}\ }\textbf {\bibinfo {volume} {09}},\ \bibinfo {pages} {016} (\bibinfo {year} {2015})},\ \Eprint {http://arxiv.org/abs/1503.06209} {arXiv:1503.06209 [astro-ph.HE]} \BibitemShut {NoStop}%
\bibitem [{\citenamefont {Brown}\ \emph {et~al.}(2019)\citenamefont {Brown}, \citenamefont {Massey}, \citenamefont {Lacroix}, \citenamefont {Strigari}, \citenamefont {Fattahi},\ and\ \citenamefont {B\oe{}hm}}]{Brown:2019whs}%
  \BibitemOpen
  \bibfield  {author} {\bibinfo {author} {\bibfnamefont {A.~M.}\ \bibnamefont {Brown}}, \bibinfo {author} {\bibfnamefont {R.}~\bibnamefont {Massey}}, \bibinfo {author} {\bibfnamefont {T.}~\bibnamefont {Lacroix}}, \bibinfo {author} {\bibfnamefont {L.~E.}\ \bibnamefont {Strigari}}, \bibinfo {author} {\bibfnamefont {A.}~\bibnamefont {Fattahi}}, \ and\ \bibinfo {author} {\bibfnamefont {C.}~\bibnamefont {B\oe{}hm}},\ }\href@noop {} {\  (\bibinfo {year} {2019})},\ \Eprint {http://arxiv.org/abs/1907.08564} {arXiv:1907.08564 [astro-ph.HE]} \BibitemShut {NoStop}%
\bibitem [{\citenamefont {Reynoso-Cordova}\ \emph {et~al.}(2021)\citenamefont {Reynoso-Cordova}, \citenamefont {Burgue\~no}, \citenamefont {Geringer-Sameth}, \citenamefont {Gonzalez-Morales}, \citenamefont {Profumo},\ and\ \citenamefont {Valenzuela}}]{Reynoso-Cordova:2019biv}%
  \BibitemOpen
  \bibfield  {author} {\bibinfo {author} {\bibfnamefont {J.}~\bibnamefont {Reynoso-Cordova}}, \bibinfo {author} {\bibfnamefont {O.}~\bibnamefont {Burgue\~no}}, \bibinfo {author} {\bibfnamefont {A.}~\bibnamefont {Geringer-Sameth}}, \bibinfo {author} {\bibfnamefont {A.~X.}\ \bibnamefont {Gonzalez-Morales}}, \bibinfo {author} {\bibfnamefont {S.}~\bibnamefont {Profumo}}, \ and\ \bibinfo {author} {\bibfnamefont {O.}~\bibnamefont {Valenzuela}},\ }\href {\doibase 10.1088/1475-7516/2021/02/010} {\bibfield  {journal} {\bibinfo  {journal} {JCAP}\ }\textbf {\bibinfo {volume} {02}},\ \bibinfo {pages} {010} (\bibinfo {year} {2021})},\ \Eprint {http://arxiv.org/abs/1907.06682} {arXiv:1907.06682 [astro-ph.HE]} \BibitemShut {NoStop}%
\bibitem [{\citenamefont {Evans}\ \emph {et~al.}(2022)\citenamefont {Evans}, \citenamefont {Strigari},\ and\ \citenamefont {Zivick}}]{Evans:2021bsh}%
  \BibitemOpen
  \bibfield  {author} {\bibinfo {author} {\bibfnamefont {A.~J.}\ \bibnamefont {Evans}}, \bibinfo {author} {\bibfnamefont {L.~E.}\ \bibnamefont {Strigari}}, \ and\ \bibinfo {author} {\bibfnamefont {P.}~\bibnamefont {Zivick}},\ }\href {\doibase 10.1093/mnras/stac261} {\bibfield  {journal} {\bibinfo  {journal} {Mon. Not. Roy. Astron. Soc.}\ }\textbf {\bibinfo {volume} {511}},\ \bibinfo {pages} {4251} (\bibinfo {year} {2022})},\ \Eprint {http://arxiv.org/abs/2109.10998} {arXiv:2109.10998 [astro-ph.GA]} \BibitemShut {NoStop}%
\bibitem [{\citenamefont {{Smith}}\ \emph {et~al.}(2023)\citenamefont {{Smith}}, \citenamefont {{Cerny}}, \citenamefont {{Hayes}}, \citenamefont {{Sestito}}, \citenamefont {{Jensen}}, \citenamefont {{McConnachie}}, \citenamefont {{Geha}}, \citenamefont {{Navarro}}, \citenamefont {{Li}}, \citenamefont {{Cuillandre}}, \citenamefont {{Errani}}, \citenamefont {{Chambers}}, \citenamefont {{Gwyn}}, \citenamefont {{Hammer}}, \citenamefont {{Hudson}}, \citenamefont {{Magnier}},\ and\ \citenamefont {{Martin}}}]{2023arXiv231110147S}%
  \BibitemOpen
  \bibfield  {author} {\bibinfo {author} {\bibfnamefont {S.~E.~T.}\ \bibnamefont {{Smith}}}, \bibinfo {author} {\bibfnamefont {W.}~\bibnamefont {{Cerny}}}, \bibinfo {author} {\bibfnamefont {C.~R.}\ \bibnamefont {{Hayes}}}, \bibinfo {author} {\bibfnamefont {F.}~\bibnamefont {{Sestito}}}, \bibinfo {author} {\bibfnamefont {J.}~\bibnamefont {{Jensen}}}, \bibinfo {author} {\bibfnamefont {A.~W.}\ \bibnamefont {{McConnachie}}}, \bibinfo {author} {\bibfnamefont {M.}~\bibnamefont {{Geha}}}, \bibinfo {author} {\bibfnamefont {J.}~\bibnamefont {{Navarro}}}, \bibinfo {author} {\bibfnamefont {T.~S.}\ \bibnamefont {{Li}}}, \bibinfo {author} {\bibfnamefont {J.-C.}\ \bibnamefont {{Cuillandre}}}, \bibinfo {author} {\bibfnamefont {R.}~\bibnamefont {{Errani}}}, \bibinfo {author} {\bibfnamefont {K.}~\bibnamefont {{Chambers}}}, \bibinfo {author} {\bibfnamefont {S.}~\bibnamefont {{Gwyn}}}, \bibinfo {author} {\bibfnamefont {F.}~\bibnamefont {{Hammer}}}, \bibinfo {author} {\bibfnamefont {M.~J.}\ \bibnamefont {{Hudson}}}, \bibinfo
  {author} {\bibfnamefont {E.}~\bibnamefont {{Magnier}}}, \ and\ \bibinfo {author} {\bibfnamefont {N.}~\bibnamefont {{Martin}}},\ }\href@noop {} {\bibfield  {journal} {\bibinfo  {journal} {arXiv e-prints}\ ,\ \bibinfo {eid} {arXiv:2311.10147}} (\bibinfo {year} {2023})},\ \Eprint {http://arxiv.org/abs/2311.10147} {arXiv:2311.10147 [astro-ph.GA]} \BibitemShut {NoStop}%
\bibitem [{\citenamefont {{Pace}}\ and\ \citenamefont {{Strigari}}(2019)}]{2019MNRAS.482.3480P}%
  \BibitemOpen
  \bibfield  {author} {\bibinfo {author} {\bibfnamefont {A.~B.}\ \bibnamefont {{Pace}}}\ and\ \bibinfo {author} {\bibfnamefont {L.~E.}\ \bibnamefont {{Strigari}}},\ }\href {\doibase 10.1093/mnras/sty2839} {\bibfield  {journal} {\bibinfo  {journal} {\mnras}\ }\textbf {\bibinfo {volume} {482}},\ \bibinfo {pages} {3480} (\bibinfo {year} {2019})},\ \Eprint {http://arxiv.org/abs/1802.06811} {arXiv:1802.06811 [astro-ph.GA]} \BibitemShut {NoStop}%
\bibitem [{\citenamefont {{Errani}}\ \emph {et~al.}(2023)\citenamefont {{Errani}}, \citenamefont {{Navarro}}, \citenamefont {{Smith}},\ and\ \citenamefont {{McConnachie}}}]{2023arXiv231110134E}%
  \BibitemOpen
  \bibfield  {author} {\bibinfo {author} {\bibfnamefont {R.}~\bibnamefont {{Errani}}}, \bibinfo {author} {\bibfnamefont {J.~F.}\ \bibnamefont {{Navarro}}}, \bibinfo {author} {\bibfnamefont {S.~E.~T.}\ \bibnamefont {{Smith}}}, \ and\ \bibinfo {author} {\bibfnamefont {A.~W.}\ \bibnamefont {{McConnachie}}},\ }\href@noop {} {\bibfield  {journal} {\bibinfo  {journal} {arXiv e-prints}\ ,\ \bibinfo {eid} {arXiv:2311.10134}} (\bibinfo {year} {2023})},\ \Eprint {http://arxiv.org/abs/2311.10134} {arXiv:2311.10134 [astro-ph.GA]} \BibitemShut {NoStop}%
\bibitem [{\citenamefont {Calore}\ \emph {et~al.}(2015)\citenamefont {Calore}, \citenamefont {Cholis},\ and\ \citenamefont {Weniger}}]{Calore:2014xka}%
  \BibitemOpen
  \bibfield  {author} {\bibinfo {author} {\bibfnamefont {F.}~\bibnamefont {Calore}}, \bibinfo {author} {\bibfnamefont {I.}~\bibnamefont {Cholis}}, \ and\ \bibinfo {author} {\bibfnamefont {C.}~\bibnamefont {Weniger}},\ }\href {\doibase 10.1088/1475-7516/2015/03/038} {\bibfield  {journal} {\bibinfo  {journal} {JCAP}\ }\textbf {\bibinfo {volume} {03}},\ \bibinfo {pages} {038} (\bibinfo {year} {2015})},\ \Eprint {http://arxiv.org/abs/1409.0042} {arXiv:1409.0042 [astro-ph.CO]} \BibitemShut {NoStop}%
\bibitem [{\citenamefont {Abdollahi}\ \emph {et~al.}(2020)\citenamefont {Abdollahi} \emph {et~al.}}]{Fermi-LAT:2019yla}%
  \BibitemOpen
  \bibfield  {author} {\bibinfo {author} {\bibfnamefont {S.}~\bibnamefont {Abdollahi}} \emph {et~al.} (\bibinfo {collaboration} {Fermi-LAT}),\ }\href {\doibase 10.3847/1538-4365/ab6bcb} {\bibfield  {journal} {\bibinfo  {journal} {Astrophys. J. Suppl.}\ }\textbf {\bibinfo {volume} {247}},\ \bibinfo {pages} {33} (\bibinfo {year} {2020})},\ \Eprint {http://arxiv.org/abs/1902.10045} {arXiv:1902.10045 [astro-ph.HE]} \BibitemShut {NoStop}%
\bibitem [{\citenamefont {Daylan}\ \emph {et~al.}(2016)\citenamefont {Daylan}, \citenamefont {Finkbeiner}, \citenamefont {Hooper}, \citenamefont {Linden}, \citenamefont {Portillo}, \citenamefont {Rodd},\ and\ \citenamefont {Slatyer}}]{Daylan:2014rsa}%
  \BibitemOpen
  \bibfield  {author} {\bibinfo {author} {\bibfnamefont {T.}~\bibnamefont {Daylan}}, \bibinfo {author} {\bibfnamefont {D.~P.}\ \bibnamefont {Finkbeiner}}, \bibinfo {author} {\bibfnamefont {D.}~\bibnamefont {Hooper}}, \bibinfo {author} {\bibfnamefont {T.}~\bibnamefont {Linden}}, \bibinfo {author} {\bibfnamefont {S.~K.~N.}\ \bibnamefont {Portillo}}, \bibinfo {author} {\bibfnamefont {N.~L.}\ \bibnamefont {Rodd}}, \ and\ \bibinfo {author} {\bibfnamefont {T.~R.}\ \bibnamefont {Slatyer}},\ }\href {\doibase 10.1016/j.dark.2015.12.005} {\bibfield  {journal} {\bibinfo  {journal} {Phys. Dark Univ.}\ }\textbf {\bibinfo {volume} {12}},\ \bibinfo {pages} {1} (\bibinfo {year} {2016})},\ \Eprint {http://arxiv.org/abs/1402.6703} {arXiv:1402.6703 [astro-ph.HE]} \BibitemShut {NoStop}%
\bibitem [{\citenamefont {Gondolo}\ \emph {et~al.}(2004)\citenamefont {Gondolo}, \citenamefont {Edsjo}, \citenamefont {Ullio}, \citenamefont {Bergstrom}, \citenamefont {Schelke},\ and\ \citenamefont {Baltz}}]{Gondolo:2004sc}%
  \BibitemOpen
  \bibfield  {author} {\bibinfo {author} {\bibfnamefont {P.}~\bibnamefont {Gondolo}}, \bibinfo {author} {\bibfnamefont {J.}~\bibnamefont {Edsjo}}, \bibinfo {author} {\bibfnamefont {P.}~\bibnamefont {Ullio}}, \bibinfo {author} {\bibfnamefont {L.}~\bibnamefont {Bergstrom}}, \bibinfo {author} {\bibfnamefont {M.}~\bibnamefont {Schelke}}, \ and\ \bibinfo {author} {\bibfnamefont {E.~A.}\ \bibnamefont {Baltz}},\ }\href {\doibase 10.1088/1475-7516/2004/07/008} {\bibfield  {journal} {\bibinfo  {journal} {JCAP}\ }\textbf {\bibinfo {volume} {07}},\ \bibinfo {pages} {008} (\bibinfo {year} {2004})},\ \Eprint {http://arxiv.org/abs/astro-ph/0406204} {arXiv:astro-ph/0406204} \BibitemShut {NoStop}%
\bibitem [{\citenamefont {{Simon}}\ and\ \citenamefont {{Geha}}(2007)}]{2007ApJ...670..313S}%
  \BibitemOpen
  \bibfield  {author} {\bibinfo {author} {\bibfnamefont {J.~D.}\ \bibnamefont {{Simon}}}\ and\ \bibinfo {author} {\bibfnamefont {M.}~\bibnamefont {{Geha}}},\ }\href {\doibase 10.1086/521816} {\bibfield  {journal} {\bibinfo  {journal} {\apj}\ }\textbf {\bibinfo {volume} {670}},\ \bibinfo {pages} {313} (\bibinfo {year} {2007})},\ \Eprint {http://arxiv.org/abs/0706.0516} {arXiv:0706.0516 [astro-ph]} \BibitemShut {NoStop}%
\bibitem [{\citenamefont {{Atwood}}\ \emph {et~al.}(2009)\citenamefont {{Atwood}}, \citenamefont {{Abdo}}, \citenamefont {{Ackermann}}, \citenamefont {{Althouse}}, \citenamefont {{Anderson}}, \citenamefont {{Axelsson}}, \citenamefont {{Baldini}}, \citenamefont {{Ballet}},\ and\ \citenamefont {{et al.}}}]{LATinstrument:2009}%
  \BibitemOpen
  \bibfield  {author} {\bibinfo {author} {\bibfnamefont {W.~B.}\ \bibnamefont {{Atwood}}}, \bibinfo {author} {\bibfnamefont {A.~A.}\ \bibnamefont {{Abdo}}}, \bibinfo {author} {\bibfnamefont {M.}~\bibnamefont {{Ackermann}}}, \bibinfo {author} {\bibfnamefont {W.}~\bibnamefont {{Althouse}}}, \bibinfo {author} {\bibfnamefont {B.}~\bibnamefont {{Anderson}}}, \bibinfo {author} {\bibfnamefont {M.}~\bibnamefont {{Axelsson}}}, \bibinfo {author} {\bibfnamefont {L.}~\bibnamefont {{Baldini}}}, \bibinfo {author} {\bibfnamefont {J.}~\bibnamefont {{Ballet}}}, \ and\ \bibinfo {author} {\bibnamefont {{et al.}}},\ }\href {\doibase 10.1088/0004-637X/697/2/1071} {\bibfield  {journal} {\bibinfo  {journal} {\apj}\ }\textbf {\bibinfo {volume} {697}},\ \bibinfo {pages} {1071} (\bibinfo {year} {2009})},\ \Eprint {http://arxiv.org/abs/0902.1089} {arXiv:0902.1089 [astro-ph.IM]} \BibitemShut {NoStop}%
\bibitem [{\citenamefont {{Wood}}\ \emph {et~al.}(2017)\citenamefont {{Wood}}, \citenamefont {{Caputo}}, \citenamefont {{Charles}}, \citenamefont {{Di Mauro}}, \citenamefont {{Magill}}, \citenamefont {{Perkins}},\ and\ \citenamefont {{Fermi-LAT Collaboration}}}]{fermipy:2017}%
  \BibitemOpen
  \bibfield  {author} {\bibinfo {author} {\bibfnamefont {M.}~\bibnamefont {{Wood}}}, \bibinfo {author} {\bibfnamefont {R.}~\bibnamefont {{Caputo}}}, \bibinfo {author} {\bibfnamefont {E.}~\bibnamefont {{Charles}}}, \bibinfo {author} {\bibfnamefont {M.}~\bibnamefont {{Di Mauro}}}, \bibinfo {author} {\bibfnamefont {J.}~\bibnamefont {{Magill}}}, \bibinfo {author} {\bibfnamefont {J.~S.}\ \bibnamefont {{Perkins}}}, \ and\ \bibinfo {author} {\bibnamefont {{Fermi-LAT Collaboration}}},\ }in\ \href {\doibase 10.22323/1.301.0824} {\emph {\bibinfo {booktitle} {35th International Cosmic Ray Conference (ICRC2017)}}},\ \bibinfo {series} {International Cosmic Ray Conference}, Vol.\ \bibinfo {volume} {301}\ (\bibinfo {year} {2017})\ p.\ \bibinfo {pages} {824},\ \Eprint {http://arxiv.org/abs/1707.09551} {arXiv:1707.09551 [astro-ph.IM]} \BibitemShut {NoStop}%
\bibitem [{\citenamefont {{Abdollahi}}\ \emph {et~al.}(2022)\citenamefont {{Abdollahi}}, \citenamefont {{Acero}}, \citenamefont {{Baldini}}, \citenamefont {{Ballet}}, \citenamefont {{Bastieri}}, \citenamefont {{Bellazzini}}, \citenamefont {{Berenji}}, \citenamefont {{Berretta}},\ and\ \citenamefont {{et al.}}}]{4FGL-DR3:2022}%
  \BibitemOpen
  \bibfield  {author} {\bibinfo {author} {\bibfnamefont {S.}~\bibnamefont {{Abdollahi}}}, \bibinfo {author} {\bibfnamefont {F.}~\bibnamefont {{Acero}}}, \bibinfo {author} {\bibfnamefont {L.}~\bibnamefont {{Baldini}}}, \bibinfo {author} {\bibfnamefont {J.}~\bibnamefont {{Ballet}}}, \bibinfo {author} {\bibfnamefont {D.}~\bibnamefont {{Bastieri}}}, \bibinfo {author} {\bibfnamefont {R.}~\bibnamefont {{Bellazzini}}}, \bibinfo {author} {\bibfnamefont {B.}~\bibnamefont {{Berenji}}}, \bibinfo {author} {\bibfnamefont {A.}~\bibnamefont {{Berretta}}}, \ and\ \bibinfo {author} {\bibnamefont {{et al.}}},\ }\href {\doibase 10.3847/1538-4365/ac6751} {\bibfield  {journal} {\bibinfo  {journal} {\apjs}\ }\textbf {\bibinfo {volume} {260}},\ \bibinfo {eid} {53} (\bibinfo {year} {2022})},\ \Eprint {http://arxiv.org/abs/2201.11184} {arXiv:2201.11184 [astro-ph.HE]} \BibitemShut {NoStop}%
\bibitem [{\citenamefont {Wilks}(1938)}]{wilks1938}%
  \BibitemOpen
  \bibfield  {author} {\bibinfo {author} {\bibfnamefont {S.~S.}\ \bibnamefont {Wilks}},\ }\href {\doibase 10.1214/aoms/1177732360} {\bibfield  {journal} {\bibinfo  {journal} {Ann. Math. Statist.}\ }\textbf {\bibinfo {volume} {9}},\ \bibinfo {pages} {60} (\bibinfo {year} {1938})}\BibitemShut {NoStop}%
\bibitem [{\citenamefont {{Di Mauro}}\ \emph {et~al.}(2022)\citenamefont {{Di Mauro}}, \citenamefont {{Stref}},\ and\ \citenamefont {{Calore}}}]{2022PhRvD.106l3032D}%
  \BibitemOpen
  \bibfield  {author} {\bibinfo {author} {\bibfnamefont {M.}~\bibnamefont {{Di Mauro}}}, \bibinfo {author} {\bibfnamefont {M.}~\bibnamefont {{Stref}}}, \ and\ \bibinfo {author} {\bibfnamefont {F.}~\bibnamefont {{Calore}}},\ }\href {\doibase 10.1103/PhysRevD.106.123032} {\bibfield  {journal} {\bibinfo  {journal} {\prd}\ }\textbf {\bibinfo {volume} {106}},\ \bibinfo {eid} {123032} (\bibinfo {year} {2022})},\ \Eprint {http://arxiv.org/abs/2212.06850} {arXiv:2212.06850 [astro-ph.HE]} \BibitemShut {NoStop}%
\bibitem [{\citenamefont {{Amorisco}}\ and\ \citenamefont {{Evans}}(2011)}]{2011MNRAS.411.2118A}%
  \BibitemOpen
  \bibfield  {author} {\bibinfo {author} {\bibfnamefont {N.~C.}\ \bibnamefont {{Amorisco}}}\ and\ \bibinfo {author} {\bibfnamefont {N.~W.}\ \bibnamefont {{Evans}}},\ }\href {\doibase 10.1111/j.1365-2966.2010.17715.x} {\bibfield  {journal} {\bibinfo  {journal} {\mnras}\ }\textbf {\bibinfo {volume} {411}},\ \bibinfo {pages} {2118} (\bibinfo {year} {2011})},\ \Eprint {http://arxiv.org/abs/1009.1813} {arXiv:1009.1813 [astro-ph.GA]} \BibitemShut {NoStop}%
\bibitem [{\citenamefont {Evans}\ \emph {et~al.}(2016)\citenamefont {Evans}, \citenamefont {Sanders},\ and\ \citenamefont {Geringer-Sameth}}]{Evans:2016xwx}%
  \BibitemOpen
  \bibfield  {author} {\bibinfo {author} {\bibfnamefont {N.~W.}\ \bibnamefont {Evans}}, \bibinfo {author} {\bibfnamefont {J.~L.}\ \bibnamefont {Sanders}}, \ and\ \bibinfo {author} {\bibfnamefont {A.}~\bibnamefont {Geringer-Sameth}},\ }\href {\doibase 10.1103/PhysRevD.93.103512} {\bibfield  {journal} {\bibinfo  {journal} {Phys. Rev. D}\ }\textbf {\bibinfo {volume} {93}},\ \bibinfo {pages} {103512} (\bibinfo {year} {2016})},\ \Eprint {http://arxiv.org/abs/1604.05599} {arXiv:1604.05599 [astro-ph.GA]} \BibitemShut {NoStop}%
\bibitem [{\citenamefont {{Navarro}}\ \emph {et~al.}(1996)\citenamefont {{Navarro}}, \citenamefont {{Frenk}},\ and\ \citenamefont {{White}}}]{1996ApJ...462..563N}%
  \BibitemOpen
  \bibfield  {author} {\bibinfo {author} {\bibfnamefont {J.~F.}\ \bibnamefont {{Navarro}}}, \bibinfo {author} {\bibfnamefont {C.~S.}\ \bibnamefont {{Frenk}}}, \ and\ \bibinfo {author} {\bibfnamefont {S.~D.~M.}\ \bibnamefont {{White}}},\ }\href {\doibase 10.1086/177173} {\bibfield  {journal} {\bibinfo  {journal} {\apj}\ }\textbf {\bibinfo {volume} {462}},\ \bibinfo {pages} {563} (\bibinfo {year} {1996})},\ \Eprint {http://arxiv.org/abs/astro-ph/9508025} {arXiv:astro-ph/9508025 [astro-ph]} \BibitemShut {NoStop}%
\bibitem [{\citenamefont {{Kuhlen}}\ \emph {et~al.}(2008)\citenamefont {{Kuhlen}}, \citenamefont {{Diemand}},\ and\ \citenamefont {{Madau}}}]{2008ApJ...686..262K}%
  \BibitemOpen
  \bibfield  {author} {\bibinfo {author} {\bibfnamefont {M.}~\bibnamefont {{Kuhlen}}}, \bibinfo {author} {\bibfnamefont {J.}~\bibnamefont {{Diemand}}}, \ and\ \bibinfo {author} {\bibfnamefont {P.}~\bibnamefont {{Madau}}},\ }\href {\doibase 10.1086/590337} {\bibfield  {journal} {\bibinfo  {journal} {\apj}\ }\textbf {\bibinfo {volume} {686}},\ \bibinfo {pages} {262} (\bibinfo {year} {2008})},\ \Eprint {http://arxiv.org/abs/0805.4416} {arXiv:0805.4416 [astro-ph]} \BibitemShut {NoStop}%
\bibitem [{\citenamefont {{Cirelli}}\ \emph {et~al.}(2011)\citenamefont {{Cirelli}}, \citenamefont {{Corcella}}, \citenamefont {{Hektor}}, \citenamefont {{H{\"u}tsi}}, \citenamefont {{Kadastik}}, \citenamefont {{Panci}}, \citenamefont {{Raidal}}, \citenamefont {{Sala}},\ and\ \citenamefont {{Strumia}}}]{Cirelli:2011}%
  \BibitemOpen
  \bibfield  {author} {\bibinfo {author} {\bibfnamefont {M.}~\bibnamefont {{Cirelli}}}, \bibinfo {author} {\bibfnamefont {G.}~\bibnamefont {{Corcella}}}, \bibinfo {author} {\bibfnamefont {A.}~\bibnamefont {{Hektor}}}, \bibinfo {author} {\bibfnamefont {G.}~\bibnamefont {{H{\"u}tsi}}}, \bibinfo {author} {\bibfnamefont {M.}~\bibnamefont {{Kadastik}}}, \bibinfo {author} {\bibfnamefont {P.}~\bibnamefont {{Panci}}}, \bibinfo {author} {\bibfnamefont {M.}~\bibnamefont {{Raidal}}}, \bibinfo {author} {\bibfnamefont {F.}~\bibnamefont {{Sala}}}, \ and\ \bibinfo {author} {\bibfnamefont {A.}~\bibnamefont {{Strumia}}},\ }\href {\doibase 10.1088/1475-7516/2011/03/051} {\bibfield  {journal} {\bibinfo  {journal} {\jcap}\ }\textbf {\bibinfo {volume} {2011}},\ \bibinfo {eid} {051} (\bibinfo {year} {2011})},\ \Eprint {http://arxiv.org/abs/1012.4515} {arXiv:1012.4515 [hep-ph]} \BibitemShut {NoStop}%
\bibitem [{\citenamefont {Drlica-Wagner}\ \emph {et~al.}(2019)\citenamefont {Drlica-Wagner} \emph {et~al.}}]{LSSTDarkMatterGroup:2019mwo}%
  \BibitemOpen
  \bibfield  {author} {\bibinfo {author} {\bibfnamefont {A.}~\bibnamefont {Drlica-Wagner}} \emph {et~al.} (\bibinfo {collaboration} {LSST Dark Matter Group}),\ }\href@noop {} {\  (\bibinfo {year} {2019})},\ \Eprint {http://arxiv.org/abs/1902.01055} {arXiv:1902.01055 [astro-ph.CO]} \BibitemShut {NoStop}%
\bibitem [{\citenamefont {Ando}\ \emph {et~al.}(2019)\citenamefont {Ando} \emph {et~al.}}]{Ando:2019rvr}%
  \BibitemOpen
  \bibfield  {author} {\bibinfo {author} {\bibfnamefont {S.}~\bibnamefont {Ando}} \emph {et~al.},\ }\href {\doibase 10.1088/1475-7516/2019/10/040} {\bibfield  {journal} {\bibinfo  {journal} {JCAP}\ }\textbf {\bibinfo {volume} {10}},\ \bibinfo {pages} {040} (\bibinfo {year} {2019})},\ \Eprint {http://arxiv.org/abs/1905.07128} {arXiv:1905.07128 [astro-ph.CO]} \BibitemShut {NoStop}%
\bibitem [{\citenamefont {{Skidmore}}\ \emph {et~al.}(2015)\citenamefont {{Skidmore}}, \citenamefont {{TMT International Science Development Teams}},\ and\ \citenamefont {{Science Advisory Committee}}}]{2015RAA....15.1945S}%
  \BibitemOpen
  \bibfield  {author} {\bibinfo {author} {\bibfnamefont {W.}~\bibnamefont {{Skidmore}}}, \bibinfo {author} {\bibnamefont {{TMT International Science Development Teams}}}, \ and\ \bibinfo {author} {\bibfnamefont {T.}~\bibnamefont {{Science Advisory Committee}}},\ }\href {\doibase 10.1088/1674-4527/15/12/001} {\bibfield  {journal} {\bibinfo  {journal} {Research in Astronomy and Astrophysics}\ }\textbf {\bibinfo {volume} {15}},\ \bibinfo {eid} {1945} (\bibinfo {year} {2015})},\ \Eprint {http://arxiv.org/abs/1505.01195} {arXiv:1505.01195 [astro-ph.IM]} \BibitemShut {NoStop}%
\bibitem [{\citenamefont {{McConnachie}}\ \emph {et~al.}(2016)\citenamefont {{McConnachie}}, \citenamefont {{Babusiaux}}, \citenamefont {{Balogh}}, \citenamefont {{Caffau}}, \citenamefont {{C{\^o}t{\'e}}}, \citenamefont {{Driver}}, \citenamefont {{Robotham}}, \citenamefont {{Starkenburg}}, \citenamefont {{Venn}}, \citenamefont {{Walker}}, \citenamefont {{Bauman}}, \citenamefont {{Flagey}}, \citenamefont {{Ho}}, \citenamefont {{Isani}}, \citenamefont {{Laychak}}, \citenamefont {{Mignot}}, \citenamefont {{Murowinski}}, \citenamefont {{Salmon}}, \citenamefont {{Simons}}, \citenamefont {{Szeto}}, \citenamefont {{Vermeulen}},\ and\ \citenamefont {{Withington}}}]{2016arXiv160600060M}%
  \BibitemOpen
  \bibfield  {author} {\bibinfo {author} {\bibfnamefont {A.~W.}\ \bibnamefont {{McConnachie}}}, \bibinfo {author} {\bibfnamefont {C.}~\bibnamefont {{Babusiaux}}}, \bibinfo {author} {\bibfnamefont {M.}~\bibnamefont {{Balogh}}}, \bibinfo {author} {\bibfnamefont {E.}~\bibnamefont {{Caffau}}}, \bibinfo {author} {\bibfnamefont {P.}~\bibnamefont {{C{\^o}t{\'e}}}}, \bibinfo {author} {\bibfnamefont {S.}~\bibnamefont {{Driver}}}, \bibinfo {author} {\bibfnamefont {A.}~\bibnamefont {{Robotham}}}, \bibinfo {author} {\bibfnamefont {E.}~\bibnamefont {{Starkenburg}}}, \bibinfo {author} {\bibfnamefont {K.}~\bibnamefont {{Venn}}}, \bibinfo {author} {\bibfnamefont {M.}~\bibnamefont {{Walker}}}, \bibinfo {author} {\bibfnamefont {S.~E.}\ \bibnamefont {{Bauman}}}, \bibinfo {author} {\bibfnamefont {N.}~\bibnamefont {{Flagey}}}, \bibinfo {author} {\bibfnamefont {K.}~\bibnamefont {{Ho}}}, \bibinfo {author} {\bibfnamefont {S.}~\bibnamefont {{Isani}}}, \bibinfo {author} {\bibfnamefont {M.~B.}\ \bibnamefont {{Laychak}}}, \bibinfo
  {author} {\bibfnamefont {S.}~\bibnamefont {{Mignot}}}, \bibinfo {author} {\bibfnamefont {R.}~\bibnamefont {{Murowinski}}}, \bibinfo {author} {\bibfnamefont {D.}~\bibnamefont {{Salmon}}}, \bibinfo {author} {\bibfnamefont {D.}~\bibnamefont {{Simons}}}, \bibinfo {author} {\bibfnamefont {K.}~\bibnamefont {{Szeto}}}, \bibinfo {author} {\bibfnamefont {T.}~\bibnamefont {{Vermeulen}}}, \ and\ \bibinfo {author} {\bibfnamefont {K.}~\bibnamefont {{Withington}}},\ }\href {\doibase 10.48550/arXiv.1606.00060} {\bibfield  {journal} {\bibinfo  {journal} {arXiv e-prints}\ ,\ \bibinfo {eid} {arXiv:1606.00060}} (\bibinfo {year} {2016})},\ \Eprint {http://arxiv.org/abs/1606.00060} {arXiv:1606.00060 [astro-ph.IM]} \BibitemShut {NoStop}%
\bibitem [{\citenamefont {{Fanson}}\ \emph {et~al.}(2018)\citenamefont {{Fanson}}, \citenamefont {{McCarthy}}, \citenamefont {{Bernstein}}, \citenamefont {{Angeli}}, \citenamefont {{Ashby}}, \citenamefont {{Bigelow}}, \citenamefont {{Bouchez}}, \citenamefont {{Burgett}}, \citenamefont {{Chauvin}}, \citenamefont {{Contos}}, \citenamefont {{Figueroa}}, \citenamefont {{Gray}}, \citenamefont {{Groark}}, \citenamefont {{Laskin}}, \citenamefont {{Millan-Gabet}}, \citenamefont {{Rakich}}, \citenamefont {{Sandoval}}, \citenamefont {{Pi}},\ and\ \citenamefont {{Wheeler}}}]{2018SPIE10700E..12F}%
  \BibitemOpen
  \bibfield  {author} {\bibinfo {author} {\bibfnamefont {J.}~\bibnamefont {{Fanson}}}, \bibinfo {author} {\bibfnamefont {P.~J.}\ \bibnamefont {{McCarthy}}}, \bibinfo {author} {\bibfnamefont {R.}~\bibnamefont {{Bernstein}}}, \bibinfo {author} {\bibfnamefont {G.}~\bibnamefont {{Angeli}}}, \bibinfo {author} {\bibfnamefont {D.}~\bibnamefont {{Ashby}}}, \bibinfo {author} {\bibfnamefont {B.}~\bibnamefont {{Bigelow}}}, \bibinfo {author} {\bibfnamefont {A.}~\bibnamefont {{Bouchez}}}, \bibinfo {author} {\bibfnamefont {W.}~\bibnamefont {{Burgett}}}, \bibinfo {author} {\bibfnamefont {E.}~\bibnamefont {{Chauvin}}}, \bibinfo {author} {\bibfnamefont {A.}~\bibnamefont {{Contos}}}, \bibinfo {author} {\bibfnamefont {F.}~\bibnamefont {{Figueroa}}}, \bibinfo {author} {\bibfnamefont {P.}~\bibnamefont {{Gray}}}, \bibinfo {author} {\bibfnamefont {F.}~\bibnamefont {{Groark}}}, \bibinfo {author} {\bibfnamefont {R.}~\bibnamefont {{Laskin}}}, \bibinfo {author} {\bibfnamefont {R.}~\bibnamefont {{Millan-Gabet}}}, \bibinfo {author}
  {\bibfnamefont {A.}~\bibnamefont {{Rakich}}}, \bibinfo {author} {\bibfnamefont {R.}~\bibnamefont {{Sandoval}}}, \bibinfo {author} {\bibfnamefont {M.}~\bibnamefont {{Pi}}}, \ and\ \bibinfo {author} {\bibfnamefont {N.}~\bibnamefont {{Wheeler}}},\ }in\ \href {\doibase 10.1117/12.2313340} {\emph {\bibinfo {booktitle} {Ground-based and Airborne Telescopes VII}}},\ \bibinfo {series} {Society of Photo-Optical Instrumentation Engineers (SPIE) Conference Series}, Vol.\ \bibinfo {volume} {10700},\ \bibinfo {editor} {edited by\ \bibinfo {editor} {\bibfnamefont {H.~K.}\ \bibnamefont {{Marshall}}}\ and\ \bibinfo {editor} {\bibfnamefont {J.}~\bibnamefont {{Spyromilio}}}}\ (\bibinfo {year} {2018})\ p.\ \bibinfo {pages} {1070012}\BibitemShut {NoStop}%
\bibitem [{\citenamefont {{Astropy Collaboration}}(2022)}]{astropy:2022}%
  \BibitemOpen
  \bibfield  {author} {\bibinfo {author} {\bibnamefont {{Astropy Collaboration}}},\ }\href {\doibase 10.3847/1538-4357/ac7c74} {\bibfield  {journal} {\bibinfo  {journal} {\apj}\ }\textbf {\bibinfo {volume} {935}},\ \bibinfo {eid} {167} (\bibinfo {year} {2022})},\ \Eprint {http://arxiv.org/abs/2206.14220} {arXiv:2206.14220 [astro-ph.IM]} \BibitemShut {NoStop}%
\bibitem [{\citenamefont {Harris}\ \emph {et~al.}(2020)\citenamefont {Harris}, \citenamefont {Millman}, \citenamefont {van~der Walt}, \citenamefont {Gommers}, \citenamefont {Virtanen}, \citenamefont {Cournapeau}, \citenamefont {Wieser}, \citenamefont {Taylor}, \citenamefont {Berg}, \citenamefont {Smith}, \citenamefont {Kern}, \citenamefont {Picus}, \citenamefont {Hoyer}, \citenamefont {van Kerkwijk}, \citenamefont {Brett}, \citenamefont {Haldane}, \citenamefont {del R{\'{i}}o}, \citenamefont {Wiebe}, \citenamefont {Peterson}, \citenamefont {G{\'{e}}rard-Marchant}, \citenamefont {Sheppard}, \citenamefont {Reddy}, \citenamefont {Weckesser}, \citenamefont {Abbasi}, \citenamefont {Gohlke},\ and\ \citenamefont {Oliphant}}]{numpy:2020}%
  \BibitemOpen
  \bibfield  {author} {\bibinfo {author} {\bibfnamefont {C.~R.}\ \bibnamefont {Harris}}, \bibinfo {author} {\bibfnamefont {K.~J.}\ \bibnamefont {Millman}}, \bibinfo {author} {\bibfnamefont {S.~J.}\ \bibnamefont {van~der Walt}}, \bibinfo {author} {\bibfnamefont {R.}~\bibnamefont {Gommers}}, \bibinfo {author} {\bibfnamefont {P.}~\bibnamefont {Virtanen}}, \bibinfo {author} {\bibfnamefont {D.}~\bibnamefont {Cournapeau}}, \bibinfo {author} {\bibfnamefont {E.}~\bibnamefont {Wieser}}, \bibinfo {author} {\bibfnamefont {J.}~\bibnamefont {Taylor}}, \bibinfo {author} {\bibfnamefont {S.}~\bibnamefont {Berg}}, \bibinfo {author} {\bibfnamefont {N.~J.}\ \bibnamefont {Smith}}, \bibinfo {author} {\bibfnamefont {R.}~\bibnamefont {Kern}}, \bibinfo {author} {\bibfnamefont {M.}~\bibnamefont {Picus}}, \bibinfo {author} {\bibfnamefont {S.}~\bibnamefont {Hoyer}}, \bibinfo {author} {\bibfnamefont {M.~H.}\ \bibnamefont {van Kerkwijk}}, \bibinfo {author} {\bibfnamefont {M.}~\bibnamefont {Brett}}, \bibinfo {author} {\bibfnamefont
  {A.}~\bibnamefont {Haldane}}, \bibinfo {author} {\bibfnamefont {J.~F.}\ \bibnamefont {del R{\'{i}}o}}, \bibinfo {author} {\bibfnamefont {M.}~\bibnamefont {Wiebe}}, \bibinfo {author} {\bibfnamefont {P.}~\bibnamefont {Peterson}}, \bibinfo {author} {\bibfnamefont {P.}~\bibnamefont {G{\'{e}}rard-Marchant}}, \bibinfo {author} {\bibfnamefont {K.}~\bibnamefont {Sheppard}}, \bibinfo {author} {\bibfnamefont {T.}~\bibnamefont {Reddy}}, \bibinfo {author} {\bibfnamefont {W.}~\bibnamefont {Weckesser}}, \bibinfo {author} {\bibfnamefont {H.}~\bibnamefont {Abbasi}}, \bibinfo {author} {\bibfnamefont {C.}~\bibnamefont {Gohlke}}, \ and\ \bibinfo {author} {\bibfnamefont {T.~E.}\ \bibnamefont {Oliphant}},\ }\href {\doibase 10.1038/s41586-020-2649-2} {\bibfield  {journal} {\bibinfo  {journal} {Nature}\ }\textbf {\bibinfo {volume} {585}},\ \bibinfo {pages} {357} (\bibinfo {year} {2020})}\BibitemShut {NoStop}%
\bibitem [{\citenamefont {Hunter}(2007)}]{matplotlib:2007}%
  \BibitemOpen
  \bibfield  {author} {\bibinfo {author} {\bibfnamefont {J.~D.}\ \bibnamefont {Hunter}},\ }\href {\doibase 10.1109/MCSE.2007.55} {\bibfield  {journal} {\bibinfo  {journal} {Computing in Science \& Engineering}\ }\textbf {\bibinfo {volume} {9}},\ \bibinfo {pages} {90} (\bibinfo {year} {2007})}\BibitemShut {NoStop}%
\bibitem [{\citenamefont {pandas~development team}(2020)}]{pandas:2020}%
  \BibitemOpen
  \bibfield  {author} {\bibinfo {author} {\bibfnamefont {T.}~\bibnamefont {pandas~development team}},\ }\href {\doibase 10.5281/zenodo.3509134} {\enquote {\bibinfo {title} {pandas-dev/pandas: Pandas},}\ } (\bibinfo {year} {2020})\BibitemShut {NoStop}%
\end{thebibliography}%

\end{document}